\documentclass[twocolumn]{aastex631}

\usepackage{hyperref}
\usepackage{graphicx}
\usepackage{rotating}
\usepackage[space]{grffile}
\usepackage{latexsym}
\usepackage{textcomp}
\usepackage{longtable}
\usepackage{multirow,booktabs}
\usepackage{amsfonts,amsmath,amssymb}
\usepackage{natbib}
\usepackage{url}
\usepackage{hyperref}

\hypersetup{colorlinks=true,pdfborder={0 0 0}}
\newif\iflatexml\latexmlfalse
\DeclareGraphicsExtensions{.pdf,.PDF,.png,.PNG,.jpg,.JPG,.jpeg,.JPEG}

\usepackage[utf8]{inputenc}
\usepackage[english]{babel}

\def\geqsim{\lower.73ex\hbox{$\sim$}\llap{\raise.4ex\hbox{$>$}}$\,$}

\newcommand{\dt}{{\tt desitarget}}
\newcommand{\desi}{DESI}
\newcommand{\urlstub}[1]{\url{https://data.desi.lbl.gov/public/ets/target/catalogs/#1}}
\newcommand{\urlstubsec}[1]{\url{https://data.desi.lbl.gov/public/ets/target/#1}}
\newcommand{\lsurl}[1]{\url{https://www.legacysurvey.org/#1}}

\received{{\em 2022 August 2}}
\revised{{\em 2022 November 3}}
\accepted{{\em 2022 November 7}}
\published{{\em 2023 January 11}}

\shorttitle{\desi\ Targeting}
\shortauthors{Myers et al.}

\begin{document}

\title{The Target-selection Pipeline for the Dark Energy Spectroscopic Instrument}

\correspondingauthor{Adam D.\ Myers}
\email{amyers14@uwyo.edu}

\newcounter{affilcounter}
\author{Adam D.\ Myers}
\affil{Department of Physics and Astronomy, University of Wyoming, Laramie, WY 82071, USA}

\author[0000-0002-2733-4559]{John Moustakas} 
\affil{Department of Physics and Astronomy, Siena College, 515 Loudon Road, Loudonville, NY 12211, USA}

\author[0000-0003-4162-6619]{Stephen Bailey}
\affil{Lawrence Berkeley National Laboratory, One Cyclotron Road, Berkeley, CA 94720, USA}

\author{Benjamin A.\ Weaver}
\affil{NSF's National Optical-Infrared Astronomy Research Laboratory, 950 N. Cherry Avenue, Tucson, AZ 85719, USA}

\author[0000-0001-8274-158X]{Andrew P.\ Cooper}
\affil{Institute of Astronomy and Department of Physics, National Tsing Hua University, 101 Kuang-Fu Rd. Sec. 2, Hsinchu 30013, Taiwan}

\author{Jaime E.\ Forero-Romero} 
\affil{Departamento de F\'{i}sica, Universidad de los Andes, Cra. 1 No. 18A-10, Bogot\'{a}, Colombia}


\author[0000-0003-1820-8486]{Bela Abolfathi}
\affil{Department of Physics and Astronomy, University of California, Frederick Reines Hall, Irvine, CA 92697, USA}

\author[0000-0002-5896-6313]{David M.\ Alexander}
\affil{Centre for Extragalactic Astronomy, Department of Physics, Durham University, South Road, Durham DH1 3LE, UK}

\author{David Brooks}
\affil{Department of Physics \& Astronomy, University College London, Gower Street, London, WC1E 6BT, UK}

\author[0000-0001-8996-4874]{Edmond Chaussidon}
\affil{IRFU, CEA, Université Paris-Saclay, F-91191 Gif-sur-Yvette, France}

\author{Chia-Hsun Chuang}
\affil{Department of Physics and Astronomy, The University of Utah, 115 South 1400 East, Salt Lake City, UT 84112, USA}
\affil{Kavli Institute for Particle Astrophysics and Cosmology, Stanford University, 452 Lomita Mall, Stanford, CA 94305, USA}

\author{Kyle Dawson}
\affil{Department of Physics and Astronomy, The University of Utah, 115 South 1400 East, Salt Lake City, UT 84112, USA}

\author[0000-0002-4928-4003]{Arjun Dey}
\affil{NSF's National Optical-Infrared Astronomy Research Laboratory, 950 N. Cherry Avenue, Tucson, AZ 85719, USA}

\author[0000-0002-5665-7912]{Biprateep Dey}
\affiliation{Department of Physics and Astronomy and PITT-PACC, University of Pittsburgh, 100 Allen Hall, 3941 O'Hara St., Pittsburgh, PA 15260, USA}

\author[0000-0002-5402-1216]{Govinda Dhungana}
\affil{Department of Physics, Southern Methodist University, 3215 Daniel Avenue, Dallas, TX 75205, USA}

\author{Peter Doel}
\affil{Department of Physics \& Astronomy, University College London, Gower Street, London, WC1E 6BT, UK}

\author{Kevin Fanning}
\affil{Department of Physics, The Ohio State University, 191 West Woodruff Avenue, Columbus, OH 43210, USA}
\affil{Center for Cosmology and AstroParticle Physics, The Ohio State University, Columbus, OH 43210, USA}

\author{Enrique Gazta\~{n}aga}
\affil{Institut de C\`{i}encies de l'Espai, IEEC-CSIC, Campus UAB, Carrer de Can Magrans s/n, 08913 Bellaterra, Barcelona, Spain}

\author{Satya Gontcho A Gontcho}
\affil{Lawrence Berkeley National Laboratory, One Cyclotron Road, Berkeley, CA 94720, USA}
\affil{Department of Physics and Astronomy, University of Rochester, 500 Joseph C. Wilson Boulevard, Rochester, NY 14627, USA}

\author[0000-0003-4089-6924]{Alma X. Gonzalez-Morales}
\affil{Consejo Nacional de Ciencia y Tecnolog\'ia, Av. Insurgentes Sur 1582. Colonia Credito Constructor, Del. Benito Jurez C.P. 03940, M\'exico D.F. M\'exico}
\affil{Departamento de F\'{i}sica, Universidad de Guanajuato - DCI, C.P. 37150, Leon, Guanajuato, M\'{e}xico}

\author[0000-0003-1197-0902]{ChangHoon Hahn}
\affil{Lawrence Berkeley National Laboratory, One Cyclotron Road, Berkeley, CA 94720, USA}
\affil{Department of Astrophysical Sciences, Princeton University, Peyton Hall, Princeton NJ 08544, USA}

\author[0000-0002-9136-9609]{Hiram K.\ Herrera-Alcantar}
\affil{Departamento de F\'{i}sica, Universidad de Guanajuato - DCI, C.P. 37150, Leon, Guanajuato, M\'{e}xico}

\author{Klaus Honscheid}
\affil{Department of Physics, The Ohio State University, 191 West Woodruff Avenue, Columbus, OH 43210, USA}

\author[0000-0002-6024-466X]{Mustapha Ishak}
\affil{Department of Physics, The University of Texas at Dallas, Richardson, TX 75080, USA}

\author[0000-0002-5652-8870]{Tanveer Karim}
\affil{Harvard-Smithsonian Center for Astrophysics, 60 Garden St., Cambridge, MA 02138, USA}

\author[0000-0002-8828-5463]{David Kirkby}
\affil{Department of Physics and Astronomy, University of California, Frederick Reines Hall, Irvine, CA 92697, USA}

\author[0000-0003-3510-7134]{Theodore Kisner}
\affil{Lawrence Berkeley National Laboratory, One Cyclotron Road, Berkeley, CA 94720, USA}

\author[0000-0003-2644-135X]{Sergey E.\ Koposov}
\affiliation{Institute for Astronomy, University of Edinburgh, Royal Observatory, Blackford Hill, Edinburgh EH9 3HJ, UK}
\affiliation{Institute of Astronomy, University of Cambridge, Madingley road,  Cambridge, CB3 0HA, UK}

\author[0000-0001-6356-7424]{Anthony Kremin}
\affil{Lawrence Berkeley National Laboratory, One Cyclotron Road, Berkeley, CA 94720, USA}

\author{Ting-Wen Lan}
\affil{Department of Physics, National Taiwan University, Taipei 10617, Taiwan}
\affil{Graduate Institute of Astrophysics, National Taiwan University Taipei 10617, Taiwan}

\author[0000-0003-1838-8528]{Martin Landriau}
\affil{Lawrence Berkeley National Laboratory, One Cyclotron Road, Berkeley, CA 94720, USA}

\author[0000-0002-1172-0754]{Dustin Lang}
\affil{Perimeter Institute for Theoretical Physics, 31 Caroline St. North, Waterloo, ON N2L 2Y5, Canada}

\author[0000-0003-1887-1018]{Michael E. Levi}
\affil{Lawrence Berkeley National Laboratory, One Cyclotron Road, Berkeley, CA 94720, USA}

\author{Christophe Magneville} 
\affil{IRFU, CEA, Université Paris-Saclay, F-91191 Gif-sur-Yvette, France}

\author[0000-0002-5166-8671]{Lucas Napolitano}
\affil{Department of Physics and Astronomy, University of Wyoming, Laramie, WY 82071, USA}

\author[0000-0002-4279-4182]{Paul Martini}
\affil{Center for Cosmology and AstroParticle Physics, The Ohio State University, Columbus, OH 43210, USA} 
\affil{Department of Astronomy, The Ohio State University, Columbus, OH 43210, USA} \affil{Radcliffe Institute for Advanced Study, Harvard University, Cambridge, MA 02138, USA}

\author[0000-0002-1125-7384]{Aaron Meisner}
\affil{NSF's National Optical-Infrared Astronomy Research Laboratory, 950 N. Cherry Avenue, Tucson, AZ 85719, USA}

\author[0000-0001-8684-2222]{Jeffrey A. Newman}
\affiliation{Department of Physics and Astronomy and PITT-PACC, University of Pittsburgh, 100 Allen Hall, 3941 O'Hara St., Pittsburgh, PA 15260, USA}

\author[0000-0003-3188-784X]{Nathalie Palanque-Delabrouille}
\affil{Lawrence Berkeley National Laboratory, One Cyclotron Road, Berkeley, CA 94720, USA}
\affil{IRFU, CEA, Université Paris-Saclay, F-91191 Gif-sur-Yvette, France}

\author[0000-0002-0644-5727]{Will Percival}
\affil{Perimeter Institute for Theoretical Physics, 31 Caroline St. North, Waterloo, ON N2L 2Y5, Canada}
\affil{Department of Physics and Astronomy, University of Waterloo, 200 University Ave W, Waterloo, ON N2L 3G1, Canada}

\author{Claire Poppett}
\affiliation{Space Sciences Laboratory at University of California, 7 Gauss Way, Berkeley, CA 94720}

\author[0000-0001-7145-8674]{Francisco Prada}
\affil{Instituto de Astrofisica de Andaluc\'{i}a, Glorieta de la Astronom\'{i}a, s/n, E-18008 Granada, Spain}

\author[0000-0001-5999-7923]{Anand Raichoor}
\affil{Lawrence Berkeley National Laboratory, One Cyclotron Road, Berkeley, CA 94720, USA}

\author{Ashley J. Ross}
\affil{Center for Cosmology and AstroParticle Physics, The Ohio State University, Columbus, OH 43210, USA} 

\author[0000-0002-3569-7421]{Edward F.\ Schlafly}
\affil{Lawrence Livermore National Laboratory, P.O. Box 808 L-211, Livermore, CA 94551, USA}

\author[0000-0002-5042-5088]{David Schlegel}
\affil{Lawrence Berkeley National Laboratory, One Cyclotron Road, Berkeley, CA 94720, USA}

\author{Michael Schubnell}
\affil{Department of Physics, University of Michigan, 450 Church St, Ann Arbor, MI 48109, USA}

\author{Ting Tan}
\affil{Sorbonne Universit\'{e}, CNRS/IN2P3, Laboratoire de Physique Nucl\'{e}aire et de Hautes Energies (LPNHE), FR-75005 Paris, France}

\author[0000-0003-1704-0781]{Gregory Tarle}
\affil{Department of Physics, University of Michigan, 450 Church St, Ann Arbor, MI 48109, USA}

\author{Michael J.\ Wilson}
\affil{Lawrence Berkeley National Laboratory, One Cyclotron Road, Berkeley, CA 94720, USA}
\affil{Institute for Computational Cosmology, Department of Physics, Durham University, South Road, Durham DH1 3LE, UK}

\author{Christophe Y\`eche}
\affil{IRFU, CEA, Université Paris-Saclay, F-91191 Gif-sur-Yvette, France}

\author[0000-0001-5381-4372]{Rongpu Zhou}
\affil{Lawrence Berkeley National Laboratory, One Cyclotron Road, Berkeley, CA 94720, USA}

\author[0000-0002-4135-0977]{Zhimin Zhou}
\affil{National Astronomical Observatories, Chinese Academy of Sciences, A20 Datun Rd., Chaoyang District, Beijing, 100012, P.R. China}

\author[0000-0002-6684-3997]{Hu Zou}
\affil{National Astronomical Observatories, Chinese Academy of Sciences, A20 Datun Rd., Chaoyang District, Beijing, 100012, P.R. China}

\begin{abstract}

In 2021 May, the Dark Energy Spectroscopic Instrument (DESI) began a 5 yr survey of approximately 50 million total extragalactic and Galactic targets. The primary DESI dark-time targets are emission line galaxies (ELGs), luminous red galaxies (LRGs) and quasars (QSOs). In bright time, DESI will focus on two surveys known as the Bright Galaxy Survey (BGS) and the Milky Way Survey (MWS). DESI also observes a selection of ``secondary'' targets for bespoke science goals. This paper gives an overview of the publicly available pipeline ({\tt desitarget}) used to process targets for DESI observations. Highlights include details of the different DESI survey targeting phases, the targeting ID ({\tt TARGETID}) used to define unique targets, the bitmasks used to indicate a particular type of target, the data model and structure of DESI targeting files, and examples of how to access and use the {\tt desitarget} code base. This paper will also describe ``supporting'' DESI target classes, such as standard stars, sky locations, and random catalogs that mimic the angular selection function of DESI targets. The DESI target-selection pipeline is complex and sizable; this paper attempts to summarize the most salient information required to understand and work with DESI targeting data.

\end{abstract}



\section{Introduction}

As sky surveys have expanded in size and sophistication, the accompanying software has also become increasingly complex. Improvements in software pipelines are typically driven by, and proceed hand-in-hand with, augmentations in instrumentation and more ambitious scientific goals. Some examples of these algorithmic leaps include the development of techniques to perform spectroscopic reductions in near real-time to improve the completeness of redshift surveys \citep[e.g.][]{Ton79}, and the automated cataloging of galaxies using scanning machines in order to better sample and characterize large-scale-structure \citep[e.g.][]{Lov92}.

With the advent of multi-object spectrographs \citep[e.g.][]{Bag90, Lew02, Sme13}, spectroscopic surveys have greatly expanded in scope. This has facilitated larger campaigns with objectives that required greater accuracy and precision. Surveys with lower shot-noise are more prone to being affected by subtly heterogeneous source catalogs, and survey pipelines have had to became increasingly elaborate to mitigate inhomogeneity. Targeting for the 2dF Surveys, for example, incorporated color equations to account for different emulsions on glass plates versus films, and the survey pipeline generated masks to mitigate density fluctuations caused by bright stars, varying flux limits, satellite tracks and plate defects \citep[e.g.][]{Col01, Col03, 2QZ, Smi05}.

The Sloan Digital Sky Survey \citep[SDSS;][]{Yor00}, in particular, greatly advanced the frontiers of software development for large sky surveys. Specialist pipelines were developed for imaging reductions \citep{Lup01,Bla11}, spectroscopic reductions, source classifications and redshift-fitting \citep{Sub02,Bol12}, and tiling and fiber assignment \citep{Bla03}, among other goals. Crucially, the source code for these pipelines was made public, and the associated algorithms and data model were extensively documented \citep[e.g.][]{EDR}. 

Targeting pipelines also advanced with later iterations of the SDSS, such as the SDSS-III Baryon Oscillation Spectroscopic Survey \citep[BOSS;][]{BOSS} and the SDSS-IV extended Baryon Oscillation Spectroscopic Survey \citep[eBOSS;][]{eBOSS}. The BOSS and eBOSS quasar targeting pipelines \citep{Ros12, Mye15}, for instance, not only incorporated long-established approaches such as color cuts and cross-matches to multi-wavelength catalogs, but also implemented targeting techniques based on variability in multi-epoch data \citep{Pal11, Pal16}, forced photometry \citep{Lan14}, machine learning methods \citep{Yec10}, and rigorous Bayesian approaches \citep{Kir11, Bov11a, Bov12}.

The Dark Energy Spectroscopic Instrument (DESI) is a robotic, fiber-fed, highly multiplexed spectroscopic surveyor that operates on the Mayall 4-meter telescope at Kitt Peak National Observatory \citep{DESIinstrument, instrument}. \desi, which can obtain simultaneous spectra of almost $5000$ objects over a $\sim3^\circ$ field \citep[][T.\ Miller et al.\ 2023, in preparation]{DESIinstrument, focalplane}, is currently conducting a five-year survey of about a third of the sky. This campaign will obtain spectra for approximately 40 million galaxies and quasars \citep{Lev13, DESIscience}, which will produce about an order-of-magnitude more extragalactic redshifts than measured by BOSS and eBOSS combined \citep[see][for BOSS and eBOSS source statistics]{DR16}. DESI spectra span wavelengths of $\sim$3600--9800\,\AA\ with a blue-end resolution of $\sim$2000 growing to $\sim$5000 at the red-end. Effective exposure times are $\sim$1000\ seconds for dark-time targets and $\sim$180\ seconds in bright time \citep[see \S4 of][]{spec}. Quasar targets that DESI measures to be at redshifts of $z \geq 2.1$ are observed multiple times to improve signal-to-noise in the Lyman-$\alpha$ Forest, and can ultimately amass $\sim$4000\ seconds of exposure time (see \S4 of E.\ Schlafly et al.\ 2023, in preparation).

The sheer scale of the \desi\ experiment necessitates multiple supporting software pipelines and products. These include significant imaging from the \desi\ Legacy Imaging Surveys \citep[][D.\ Schlegel et al.\ 2023, in preparation]{BASS, Dey19}, an extensive spectroscopic reduction pipeline \citep{spec}, a template-fitting pipeline to derive classifications and redshifts for each targeted source ({\tt Redrock}; S.\ Bailey et al.\ 2023, in preparation), a pipeline to assign fibers to targets (A.\ Raichoor et al.\ 2023, in preparation), a pipeline to tile the survey and to plan and optimize observations as the campaign progresses (E.\ Schlafly et al.\ 2023, in preparation), and a pipeline to select targets for spectroscopic follow-up ({\tt desitarget}; this paper).

Target selection approaches for \desi\ are, themselves, varied and extensive. Other publications that accompany this work include a paper describing the \desi\ Survey Validation (SV) phase (DESI Collaboration et al.\ 2023, in preparation), two papers describing how visual inspection of spectra of targets acquired during SV produced truth tables to inform
target selection for the \desi\ Main Survey \citep{viqso, vigal}, and a series of five papers detailing the selection of \desi\ bright-time and dark-time science targets \citep[][see also \S\ref{sec:btanddt}]{qso, lrg, elg, mws, bgs}.

In this paper, we detail the \desi\ target selection pipeline, which we refer to throughout as \dt. In \S\ref{sec:bitmasks} we give an overview of how \desi\ uses bitmasks to record which targets were selected by each classification algorithm. In \S\ref{sec:targetid} we detail the unique identification number ({\tt TARGETID}) that \desi\ adopts to track targets. In \S\ref{sec:primary} we give a technical overview of the main \desi\ target classes. In \S\ref{sec:discussion} we highlight some important known issues and caveats that should be considered when working with \desi\ targets. Finally, we present some closing thoughts in \S\ref{sec:conclusions}. Throughout this paper, code examples are given in the {\tt Python} programming language and HEALPixel\footnote{\url{https://healpix.sourceforge.net/}.} {\tt nside} numbers \citep{HEALPix, healpy} are always expressed in the {\tt NESTED} scheme. The \dt\ codebase, and the entire history of its development, is publicly accessible\footnote{\url{https://github.com/desihub/desitarget}.}. Much of the data model for files produced by \dt, including definitions of quantities in these files, is also available online\footnote{\url{https://desidatamodel.readthedocs.io}.}.

\section{Bits and Bitmasks}
\label{sec:bitmasks}
The chief purpose of the \dt\ pipeline is to determine which sources selected by a variety of algorithms will be targeted for follow-up spectroscopy by \desi. To record that a specific source has been selected by a particular targeting algorithm, \dt\ uses a number of {\it bitmasks}. A tutorial to further help elucidate \desi\ bitmasks is available on GitHub\footnote{ \url{https://github.com/desihub/desitarget/blob/2.5.0/doc/nb/target-selection-bits-and-bitmasks.ipynb}, in Jupyter Notebook format.}.

The full set of targeting bitmasks assigned by \dt\ will be tabulated in forthcoming \desi\ Data Release papers (e.g., DESI Collaboration et al.\ 2023, in preparation). Here, though, we will introduce (and refer to) some of the principal \desi\ targeting bitmasks to help illustrate the functionality of the \dt\ pipeline.

\subsection{An Overview of \desi\ Target Classes}

The primary dark-time target classes observed by \desi\ \citep{DESIscience} are emission line galaxies \citep[ELGs;][]{ELGRN, elg}, luminous red galaxies \citep[LRGs;][]{LRGRN, lrg} and quasi-stellar objects \citep[QSOs, also known as quasars;][]{QSORN, qso}. During survey bright-time, \desi\ also targets galaxies as part of a dedicated Bright Galaxy Survey \citep[BGS;][]{BGSRN, bgs}, and a variety of Galactic sources as part of a Milky Way Survey \citep[MWS;][]{MWSRN, mws}. Finally, to calibrate the output from its spectroscopic pipeline, \desi\ also targets calibration sources such as standard stars and blank sky locations. These target classes will be described in more detail in \S\ref{sec:primary}.

The \desi\ survey also incorporates a plethora of programs that are not driven by the main cosmological goals of the primary campaign. These are referred to as ``secondary'' programs, and include, for example, studies of high-proper-motion stars in our Galaxy, a campaign to target M31 \citep{Dey22}, studies of peculiar velocities in local galaxies, broad censuses of extragalactic sources, follow-up of gravitational lenses, and identification of quasars at redshifts of $z~\geqsim~5$ (J.\ Yang et al.\ 2023, in preparation). These target classes will be detailed in forthcoming \desi\ Data Release papers (e.g., DESI Collaboration et al.\ 2023, in preparation)---but we include some technical details about \desi\ secondary targets in \S\ref{sec:sectid} and \S\ref{sec:secondary}.

All DESI primary targets, and the vast majority of secondary classes, are selected using properties from imaging. Based on algorithms applied to these photometric quantities, a {\em bitmask} (see \S\ref{sec:bits}) is constructed to designate a source as being a member of a particular target class or classes. Note that as the bits corresponding to ``QSO,'' ``ELG,'' ``LRG,'' etc. are assigned using only imaging quantities, subsequent DESI spectroscopy will reveal that some sources are inconsistent with their target class or classes, i.e., not all QSO {\em targets} will have quasar-like {\em spectra}.

\subsection{A Brief Introduction to Bitmasks}
\label{sec:bits}

As an example of how a bitmask is constructed, consider two types of source that \desi\ will target; LRGs and ELGs. Let's assume that LRGs are designated by bit 0 and ELGs by bit 1, and that the name of the variable used to store the targeting bitmask is {\tt DESI\_TARGET}. Then, {\tt DESI\_TARGET} = 1 ($\equiv 2^0$) would signify a source selected by the LRG targeting algorithm, {\tt DESI\_TARGET} = 2 ($\equiv 2^1$) would signify a source selected by the ELG targeting algorithm and {\tt DESI\_TARGET} = 3 ($\equiv 2^1 + 2^0$) would signify a source selected by {\em both} of the LRG and ELG targeting algorithms. The \dt\ pipeline uses 64-bit {\em signed} integers to represent bitmasks. But the \dt\ bit-values that are used to distinguish target classes are never negative numbers, so each of the bitmasks can register up to 63 distinct selections (bit 0 to bit 62)\footnote{The sign was reserved in case especial cases arose, see, e.g., \S\ref{sec:negativetid}.}.

A more realistic example of a DESI bitmask would be {\tt 1152921504606849827}, which equals $2^{60}+2^{11}+2^{9}+2^{8}+2^{5}+2^{1}+2^{0}$. For the DESI Main Survey (see \S\ref{sec:cmxsvmain}), a bitmask of {\tt 1152921504606849827} would signify that a target was part of the DESI Bright Galaxy Survey ($2^{60}$) as well as being selected by the LRG ($2^{0}$) and ELG ($2^{1}$) targeting algorithms---specifically, by the ELG ``low-priority" targeting algorithm ($2^{5}$). The remaining bits ($2^{11}+2^{9}+2^{8}$) are informational, and indicate that the method used to target ELGs and LRGs corresponded to the {\em northern} target selection algorithm (see \S\ref{sec:resolve}).

\subsection{Commissioning, Survey Validation, and the Main Survey}
\label{sec:cmxsvmain}

The \desi\ survey has progressed through a number of targeting phases:

\begin{itemize}
\item First, \desi\ underwent {\em commissioning} \citep[e.g.][]{Bes20, Mei20, Per20, Pop20, Sho20}. During this phase, a number of target classes were defined to calibrate the instrument, such as dither stars to test pointing models, standard stars to test throughput, and ``first light'' science targets.
The commissioning phase of \desi\ is typically abbreviated as {\tt CMX} by the \dt\ pipeline. Although a general reader is unlikely to encounter commissioning targets, it should be noted that:
\begin{itemize}
    \item {\tt CMX} observations were highly preliminary and, for expediency, were not carefully tracked and documented by \dt. From a targeting perspective, {\tt CMX} observations therefore should {\em not} be used for general scientific analyses. Nevertheless, broadly, commissioning bitmasks can be handled by replacing occurrences of {\tt SV1} or {\tt sv1} (see below) with {\tt cmx} or {\tt CMX} and occurrences of {\tt desi\_mask} with {\tt cmx\_mask}\footnote{Note that there were no specific MWS, BGS or secondary targets for {\tt CMX}, so there are no bitmasks for these program types.}.
    \item The {\tt CMX} data a general reader is most likely to encounter is a tile of observations of M33 with ${\tt TILEID} = 80615$\footnote{See, e.g., E.\ Schlafly et al.\ 2023 (in preparation) for a description of \desi\ {\tt TILEID}s.}.
\end{itemize}

\item Prior to commencing its main 5\,yr mission, \desi\ underwent a period of {\em Survey Validation} (see DESI Collaboration, 2023, in preparation). The major objectives of this  phase were to validate the end-to-end \desi\ systems, to provide observations to refine the purity and completeness of the targeting algorithms that would be used for the main \desi\ survey, and to stress-test the procedures that would be needed for day-to-day \desi\ operations. Survey Validation is typically abbreviated as {\tt SV} by the \dt\ pipeline, and SV itself was divided into three distinct stages:
\begin{itemize}
    \item {\tt SV1}: The first iteration of SV ran, mainly\footnote{A few SV1 observations were conducted alongside SV2, and occasional SV1 tiles were observed even later.}, from the night of 20201214 to the night of 20210402\footnote{The night of a \desi\ observation is recorded in the form YYYYMMDD where Y=Year, M=Month and D=Day.} and encompassed \desi\ tiles with {\tt TILEID} numbers in the interval 80605--80975\footnote{In January 2022, an additional SV1 tile was observed with {\tt TILEID}=82633.}. A total of 189 tiles were targeted in the context of SV1. These tiles were spread widely across the sky and were observed over multiple months in order to sample a wide range of observational conditions. The main goals of SV1, from a data pipeline perspective, were to collect sufficient spectra of different target classes to train target selection algorithms and to refine the spectroscopic pipeline \citep[see also][]{spec}. The targeting algorithms deployed during SV1 were deliberately loosened to allow targeting to be refined.

    \item {\tt SV2}: The second phase of SV (also referred to as ``the 0.1\% Survey'') ran from the night of 20210324 to the night of 20210511, covering \desi\ tiles in the interval 81000--81099. During SV2, 39 tiles were observed. The purpose of SV2 was to test critical end-to-end operational procedures, such as establishing the processing of ``Merged Target Ledgers'' (MTLs; see E.\ Schlafly et al.\ 2023, in preparation), in which spectroscopic classifications and redshifts are used to decide whether a target requires additional observations.

    \item {\tt SV3}: The final stage of SV (also known as ``the One-Percent Survey,'' but typically referred to as {\tt SV3} in this paper) ran from the night of 20210405 to the night of 20210610, although it was mostly completed by the end of the night of 20210513. SV3 covered \desi\ tiles with IDs in the interval 1--596. During SV3, 488 tiles were observed, 239 in dark time, 214 in bright time, and 35 as part of a ``backup'' program (see \S\ref{sec:obscon} and \S\ref{sec:backup}). Observations were conducted in 20 separate ``rosette'' patterns consisting of 11 overlapping pointings  arranged in a circle comprising a little more than $7\,{\rm deg}^2$ of total area. A few additional passes were made in some rosettes to further increase completeness. The main goals of SV3 were to conduct \desi\ observations in a mode that mimicked main-survey observations, and to sample main-survey target classes at high completeness.
\end{itemize}

The DESI collaboration typically refers to SV1 as ``Survey Validation" and SV3 as ``the One-Percent Survey" (again, see DESI Collaboration et al.\ 2023, in preparation). But, {\em this} paper will refer to the entire period encompassing SV1, SV2 and SV3 as ``Survey Validation," as such terminology better matches the nomenclature embedded in the \dt\ data model.

\item Finally, \desi\ embarked upon its five-year {\em Main Survey}, which is typically denoted {\tt main} by the \dt\ pipeline. First observations for the Main Survey began on 20210514. The target classes that had been refined using spectra from SV1 and observed at high completeness during SV3 then began to be observed in earnest. Observations were conducted using procedures that were introduced during SV2 and finalized near the end of SV3.
\end{itemize}

\begin{deluxetable*}{cc}[t]
\tablecaption{Commands for importing DESI targeting bitmasks}\label{table:bitcmds}
\tablewidth{0pt}
\tablehead{
\colhead{Phase} & 
\colhead{Import command}
}
\startdata
 {\tt SV1} & {\tt from desitarget.sv1.sv1\_targetmask import desi\_mask, bgs\_mask, mws\_mask, scnd\_mask} \tablenotemark{a} \\
 {\tt SV2} & {\tt from desitarget.sv2.sv2\_targetmask import desi\_mask, bgs\_mask, mws\_mask, scnd\_mask} \\
 {\tt SV3} & {\tt from desitarget.sv3.sv3\_targetmask import desi\_mask, bgs\_mask, mws\_mask, scnd\_mask} \\
 {\tt main} & {\tt from desitarget.targetmask import desi\_mask, bgs\_mask, mws\_mask, scnd\_mask} \\
\enddata
\tablenotetext{a}{{\tt desi\_mask}, {\tt bgs\_mask}, {\tt mws\_mask} and {\tt scnd\_mask}, here, correspond to the different target type columns listed in Table~\ref{table:bitcols}. Any of the individual masks can be omitted from the command.}
\end{deluxetable*}
\begin{deluxetable*}{ccccccc}[t]
\tablecaption{Relevant column names for DESI targeting bitmasks}\label{table:bitcols}
\tablewidth{0pt}
\tablehead{
\colhead{} & 
\colhead{} & 
\colhead{Primary Targets} & 
\colhead{} & 
\colhead{Secondary Targets} &  \\
\colhead{Survey Phase} & 
\colhead{Dark-time\tablenotemark{a}} & 
\colhead{BGS} &
\colhead{MWS} & 
\colhead{} & 
}
\startdata
 {\tt SV1} & SV1\_DESI\_TARGET & SV1\_BGS\_TARGET & SV1\_MWS\_TARGET & SV1\_SCND\_TARGET \\
 {\tt SV2} & SV2\_DESI\_TARGET & SV2\_BGS\_TARGET  & SV2\_MWS\_TARGET & SV2\_SCND\_TARGET  \\
 {\tt SV3} & SV3\_DESI\_TARGET & SV3\_BGS\_TARGET & SV3\_MWS\_TARGET & SV3\_SCND\_TARGET \\
 {\tt main} & DESI\_TARGET & BGS\_TARGET & MWS\_TARGET & SCND\_TARGET \\
\enddata
\tablenotetext{a}{Dark-time targets include LRGs, ELGs and QSOs.}
\end{deluxetable*}

\subsection{\desi\ Bitmask Denominations, Values and Access}
\label{sec:bitcols}
The \desi\ bit-names and bit-values are available on GitHub for SV1\footnote{\url{https://github.com/desihub/desitarget/blob/0.51.0/py/desitarget/sv1/data/sv1_targetmask.yaml}}, SV2\footnote{\url{https://github.com/desihub/desitarget/blob/0.53.0/py/desitarget/sv2/data/sv2_targetmask.yaml}}, SV3\footnote{\url{https://github.com/desihub/desitarget/blob/0.57.0/py/desitarget/sv3/data/sv3_targetmask.yaml}} and the Main Survey\footnote{\url{https://github.com/desihub/desitarget/blob/1.1.1/py/desitarget/data/targetmask.yaml}}. Note that,  although the vast majority of targets for SV1 used version {\tt 0.51.0} of the {\tt desitarget} code, some SV1 secondary programs were updated to use version {\tt 0.52.0} of the code (see the Appendix of DESI Collaboration et al.\ 2023, in preparation, for more details). Further, for secondary targets from SV1, it is important to use the appropriate version of the \dt\ code for the corresponding survey files (e.g.\ use version {\tt 0.51.0} of {\dt} when working with bit-masks from files in the {\tt 0.51.0/targets/sv1} directory), as some SV1 secondary targeting bits were deprecated. For all primary targets (and throughout SV2, SV3 and the Main Survey), it is acceptable to use the corresponding version of the code or a {\em later} version as bitmasks were only {\em added} during software development. The bitmasks can be accessed directly in Python using the {\tt targetmask} module in the \dt\ code. For example, if a user is interested in the primary dark-time target classes (ELGs, LRGs, QSOs, etc.) the relevant bitmask can be accessed via the command {\tt from desitarget.targetmask import desi\_mask}. The specific commands needed to retrieve each bitmask for different combinations of survey phase and target type are listed in Table~\ref{table:bitcmds}. The bits that correspond to the different bitmasks are stored in different columns depending on the type of target. For instance, the primary dark-time classes are stored in \desi\ data files in a column called {\tt DESI\_TARGET} for the Main Survey and {\tt SVX\_DESI\_TARGET} for SV for ${\tt X} \in 1, 2, 3$. The specific column names used to store each target type for each survey phase are listed in Table~\ref{table:bitcols}.

It may be helpful to illustrate the use of \desi\ targeting bitmasks with a short example. Say we wish to know if a target has been selected by (at least) the quasar targeting algorithm in the Main Survey. Further, let's assume that the appropriate data file has been read into a structure called {\tt targs}, then the necessary commands to determine which indexes in {\tt targs} correspond to quasar targets would be: \\

{\small
{\tt from desitarget.targetmask import desi\_mask as Mx} \\
\indent {\tt is\_QSO = (targs["DESI\_TARGET"] \& Mx["QSO"]) != 0} . \\  
}

\noindent Much additional functionality exists for working with \desi\ targeting bitmasks. Two utility functions, in particular, are worth mentioning. First, any integer ({\tt i}) can be converted to the corresponding bit-names using the {\tt mask.names(i)} function. For example, {\tt desi\_mask.names(7)} will return the list {\tt [`LRG', `ELG', `QSO']}. Second, the function {\tt main\_cmx\_or\_sv} can be used to directly extract the relevant bitmasks, column names, and survey phase (detailed in Table~\ref{table:bitcmds} and \ref{table:bitcols}) from a \desi\ data file. For example: \\

{\small
{\tt from desitarget.targets import main\_cmx\_or\_sv} \\
\indent {\tt column\_names, masks, survey = main\_cmx\_or\_sv(targs)} . \\  
}

\subsection{Observing Conditions and Programs}
\label{sec:obscon}
Depending on the quality of the observing conditions at Kitt Peak, \desi\ pursues one of three different observing programs, corresponding to three different sets of targets (see, e.g., E.\ Schlafly et al.\ 2023, in preparation, for more details). During {\tt DARK} time, \desi\ observes ELG, LRG and QSO targets; the bits for such targets are stored in columns with names that resemble {\tt DESI\_TARGET} (see Table~\ref{table:bitcols}) and are described in the {\tt desi\_mask} mask (see Table~\ref{table:bitcmds}). During {\tt BRIGHT} time, \desi\ observes BGS targets, stored in columns like {\tt BGS\_TARGET} and recorded in the {\tt bgs\_mask} mask, and MWS targets, stored in columns like {\tt MWS\_TARGET} and recorded in the {\tt mws\_mask} mask. Finally, during more marginal observing conditions, \desi\ pursues a {\tt BACKUP} program (see \S\ref{sec:backup}); the bits that indicate targets that are part of the backup program are also stored in columns like {\tt MWS\_TARGET} and are recorded in the {\tt mws\_mask} mask.

Because the same target can be selected to be observed in both, e.g., bright and dark time, it is useful to record the specific programs associated with each target. In \desi\ targeting files, such information is stored in a column called {\tt OBSCONDITIONS} and the corresponding bit-names and bit-values are accessible in the {\tt obsconditions} mask\footnote{\url{https://github.com/desihub/desitarget/blob/1.1.1/py/desitarget/data/targetmask.yaml\#L184-L187}.}, which can be accessed using a similar command to those listed in Table~\ref{table:bitcmds}: \\

{\small
{\tt from desitarget.targetmask import obsconditions} . \\
}

\noindent Note that there are some placeholder bits included in the {\tt obsconditions} mask (e.g., {\tt GRAY}) that were deprecated for the \desi\ Main Survey or that were never adopted for \desi\ operations.

\section{{\tt TARGETID} - A Unique Identifier for \desi\ Targets}
\label{sec:targetid}

\desi\ utilizes a unique ID, similar to the {\tt objID} used by the SDSS \citep[e.g.][]{EDR}, to track targeted objects through the end-to-end pipeline. In \desi, this unique targeting ID is denoted {\tt TARGETID}. It is critical for \desi\ operations that each {\tt TARGETID} is associated with only one object. However, it is less crucial that each {\em object} is associated with only one {\tt \textit{TARGETID}}, particularly for targets that will not be used in the key \desi\ large-scale-structure analyses. So, it is important to note that (for most cases) {\tt TARGETID} does not have coordinate-based provenance---and targets at the {\em same} coordinates can therefore have a {\em different} {\tt TARGETID}. Instead {\tt TARGETID} is constructed using information from the imaging survey used to select a target. 

An illustrative example in \desi\ might be a target selected from the Legacy Imaging Surveys versus a target selected using only information from {\em Gaia} \citep[e.g.][]{GaiaDR2}. Although an object {\em might} share a location on the sky in {\em Gaia} and the Legacy Surveys, \dt\ does not perform an exhaustive coordinate-match across all possible input surveys. Therefore, such a target can have a {\tt TARGETID} derived from Legacy Surveys information (see \S\ref{sec:standardtargetid}) {\em and} a {\tt TARGETID} derived from {\em Gaia} information (see \S\ref{sec:gaiatargetid}).

An exception to the rule that the DESI {\tt TARGETID} is not coordinate-based is {\em negative} values of {\tt TARGETID}. Negative {\tt TARGETID}s, which are described in \S\ref{sec:negativetid}, are used to distinguish sky locations that need to be rapidly assigned during survey operations.

\begin{figure*}[!t]
\begin{center}
\includegraphics[width=0.9\textwidth]{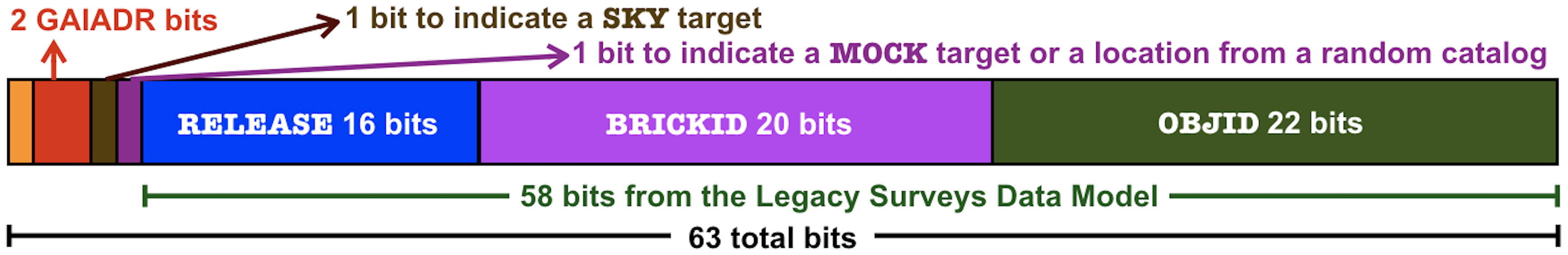}
\caption{The bits that comprise a positive \desi\ {\tt TARGETID}. The least significant bit (i.e.\ $2^0$) is depicted to the right of this diagram. The 58 least significant bits ($2^0$--$2^{57}$) record information from the Legacy Imaging Surveys. The next two most significant bits are Boolean flags that indicate whether a target is drawn from a mock or random catalog ($2^{58}$) and whether a target is a blank sky location ($2^{59}$). The remaining two populated bits ({\tt GAIADR}), which are used to indicated targets that are selected solely using {\em Gaia} information, store the {\em Gaia} Data Release number. ($2^{60}$--$2^{61}$). Bit 62 is not used. Note that DESI also uses a differently structured {\em negative} {\tt TARGETID} to distinguish some sky locations (see \S\ref{sec:negativetid}).
\label{fig:TARGETID}}
\end{center}
\end{figure*}

\subsection{Primary Targets}
\label{sec:standardtargetid}

Figure~\ref{fig:TARGETID} depicts the bitmask that defines the \desi\ {\tt TARGETID}\footnote{See also \url{https://github.com/desihub/desitarget/blob/1.1.1/py/desitarget/data/targetmask.yaml\#L214-L221}.}. In the rest of this subsection, we will detail the bits that comprise a ``primary'' {\tt TARGETID}, by which we mean the unique identifier for a source that is part of one of the primary \desi\ subprograms, i.e., LRGs, ELGs, QSOs, the BGS or the MWS, rather than being part of a secondary program (see, e.g., Table~\ref{table:bitcols}).

\subsubsection{The {\tt OBJID} and {\tt BRICKID} Bits}

Imaging from the Legacy Surveys \citep[][D. Schlegel et al.\ 2023, in preparation]{Dey19}, on which \desi\ targeting draws, is processed through the {\tt legacypipe} pipeline\footnote{\url{https://github.com/legacysurvey/legacypipe.}}
to produce catalogs of sources. The {\tt legacypipe} code extracts these sources in individual regions of the sky
denoted {\em bricks}. The bricks cover an area of roughly $0.25^\circ \times 0.25^\circ$ on the sky, and each brick is assigned a unique integer from 1 to 662{,}174.
Each extracted source in each brick is assigned an integer from 0 to $N-1$, where
$N$ is the total number of sources extracted by {\tt legacypipe} in the brick. The unique
brick number is called {\tt BRICKID} and the unique source number is called {\tt OBJID}\footnote{See, e.g., \lsurl{dr9/catalogs/}.}.

Together, {\tt OBJID} and {\tt BRICKID} encode a unique {\tt legacypipe}-extracted source. {\tt TARGETID} inherits
these two numbers as a method to track unique sources, and they occupy the two least significant sets of bits in {\tt TARGETID} (see Figure~\ref{fig:TARGETID}). 

\subsubsection{The {\tt RELEASE} Bits}

{\tt OBJID} and {\tt BRICKID} encode a unique source, but not the imaging reductions from which that source was extracted.
To track this, starting with Data Release 4 (DR4), the Legacy Surveys included a column {\tt RELEASE} that records image-processing information\footnote{See, e.g., \lsurl{release/}.}.
{\tt RELEASE} is an integer in the 1000s, with the first digit denoting the Data Release and subsequent digits representing the photometric system. 

The Legacy Imaging Surveys commenced with Data Release 1, which means {\tt RELEASE} numbers $< 1000$ have no pre-defined meaning. As we will discuss more in subsequent sections, we take advantage of this extra bit-space by using integers with {\tt RELEASE} $< 1000$ to indicate a target that (potentially) was selected using information from beyond the Legacy Surveys.

\subsubsection{The {\tt MOCK} Bit and {\tt TARGETID}s for Random Catalogs}

In preparation for going on-sky, the \desi\ collaboration produced multiple catalogs of ersatz targets to stress test the end-to-end software pipelines and to model survey outcomes. The {\tt MOCK} bit 
is a single bit (again see Figure~\ref{fig:TARGETID}) that encodes
whether or not a source is derived from such ``mock" data. In addition, the \dt\ pipeline includes a {\tt randoms} module that can be used to produce catalogs of  points that mimic the selection function of sources in the Legacy Imaging Surveys (see \S\ref{sec:randoms}). A value of ``1" for the  {\tt MOCK} bit indicates that a target was generated as part of either a ``mock'' data run or a random catalog. 

Information at locations created in \dt\ random catalogs is derived at the pixel-level using images from the Legacy Surveys (again see \S\ref{sec:randoms}). As such, {\tt RELEASE} and {\tt BRICKID} are inherited directly from the Legacy Surveys for points in random catalogs. Within a brick, \dt\ assigns an {\tt OBJID} integer for random points sequentially in order of Right Ascension (henceforth RA), such that the object with the smallest RA is assigned {\tt OBJID=0}. Ordering by RA in this manner guarantees that random points in northern and southern portions of the Legacy Surveys (see \S\ref{sec:resolve}) have the same {\tt OBJID}. Note that even though [{\tt RELEASE}, {\tt BRICKID}, {\tt OBJID}] for random points can resemble similar combinations for real sources from the Legacy Surveys, random points will always have a distinct {\tt TARGETID} because the {\tt MOCK} bit will be set.

\subsubsection{The {\tt SKY} Bit and {\tt TARGETID}s for Blank Skies}
\label{sec:skybit}

The \dt\ pipeline also generates blank sky positions to facilitate spectral calibration (see \S\ref{sec:skies}). The {\tt SKY} bit is a single bit (again see Figure~\ref{fig:TARGETID}), which, if set, indicates that a target corresponds to a sky position. 

The majority of blank sky locations in the \desi\ footprint are assigned at the pixel-level using blobmaps from the Legacy Surveys imaging (again see \S\ref{sec:skies}). For these sky locations, {\tt RELEASE} and {\tt BRICKID} are therefore inherited directly from the Legacy Surveys. Within each brick, the \dt\ code assigns sequential integers for {\tt OBJID}. This schema is guaranteed to produce unique values of {\tt TARGETID} as setting the {\tt SKY} bit distinguishes combinations of [{\tt RELEASE}, {\tt BRICKID}, {\tt OBJID}] from any target derived from the Legacy Surveys source catalogs.

\desi\ also creates files of ``supplemental'' sky targets to cover areas outside of the Legacy Imaging Surveys (see \S\ref{sec:suppskies}), which are assigned by using a catalog-level match to avoid bright sources in {\em Gaia}. If these supplemental skies {\em are} near a bright object in {\em Gaia}, they are retained but are designated to be {\tt BAD\_SKY} (see \S\ref{sec:suppskies} for more details). For {\tt BAD\_SKY} supplemental sky targets, {\tt RELEASE} is set to 0 and the {\tt GAIADR} indicative bits (see \S\ref{sec:gaiatargetid}) are set to the {\em Gaia} Data Release integer used to assign supplemental skies (typically `2' to indicate {\em Gaia} DR2). For other supplemental skies, both the {\tt RELEASE} and {\tt GAIADR} bits are populated with the {\em Gaia} DR integer. {\tt BRICKID} is set to the {\tt nside = 256} HEALPixel integer of each supplemental sky location, and {\tt OBJID} is a sequential integer generated within each HEALPixel. The {\tt SKY} bit, of course, is also always set. Note that setting {\tt RELEASE} $< 1000$ guarantees that the {\tt TARGETID} for supplemental skies will always differ from the {\tt TARGETID} for pixel-level skies, as there were no Legacy Surveys Data Releases prior to DR1.

\subsubsection{The {\tt GAIADR} bits and Gaia-based {\tt TARGETID}s}
\label{sec:gaiatargetid}

A substantial set of \desi\ targets are derived solely using information from {\em Gaia}. These targets generally lie outside of the Legacy Imaging Surveys footprint and are intended for {\tt BACKUP} observations (see \S\ref{sec:obscon}). Objects that are targeted purely using {\em Gaia} have the {\tt GAIADR} bits (depicted in Figure~\ref{fig:TARGETID}) set to the {\em Gaia} Data Release used to select the target (i.e\ `2' for {\em Gaia} DR2)\footnote{{\em Gaia} DR1 was never used for \desi\ targeting. {\tt GAIADR==1} instead denotes a ``first light'' target observed during commissioning.}. For targets that are selected solely using {\em Gaia}, {\tt BRICKID} is populated with the integer of the {\tt nside = 32} HEALPixel that contains the target, and {\tt OBJID} is set to a sequential integer generated within each HEALPixel. Note that this schema differs slightly from the one described for supplemental skies in \S\ref{sec:skybit} (for which {\tt nside = 256} is adopted  when populating {\tt BRICKID}).

\subsection{Secondary Targets}
\label{sec:sectid}

In addition to ``primary'' target classes, \desi\ observes a number of secondary programs (see also \S\ref{sec:secondary}) proposed by the wider \desi\ collaboration to address specific science goals. In many cases, proposers specified that targets should not share the observing characteristics of a \desi\ primary target. For example, a supernova in a galaxy that is part of the BGS sample should not be considered to be the same target as the galaxy itself. Such a supernova might not be at exactly the same location as the galactic center, and may, for instance, need to be scheduled for dark-time observations rather than during bright time. The end-to-end \desi\ pipeline uses {\tt TARGETID} to ensure correspondence between a target, an allocated fiber, and the resulting spectrum. Therefore, secondary targets that cannot be merged with a corresponding primary target (see \S\ref{sec:secondary}) must, by definition, have a different {\tt TARGETID}. In some instances, there is an unused primary {\tt TARGETID}, because a secondary target that {\em is} also a source from the Legacy Surveys is not part of the primary \desi\ campaign. In those instances, a secondary target is simply assigned the primary {\tt TARGETID} detailed in \S\ref{sec:standardtargetid}. Still, there are cases where a secondary target does not have a corresponding source in the Legacy Surveys {\em or} a secondary target is prevented from being merged with a primary target that is already using the primary {\tt TARGETID}. In those cases, the \dt\ pipeline constructs a ``secondary'' {\tt TARGETID} that is guaranteed to always be unique, as well as distinct from the {\tt TARGETID} of any primary target. 

\subsubsection{The General {\tt TARGETID} for Secondary Targets}
\label{sec:gensectid}

The secondary {\tt TARGETID} is structured similarly to the schema displayed in Figure~\ref{fig:TARGETID}. {\tt BRICKID} has the same definition for secondary and primary targets---it is the number of the brick from the Legacy Surveys on which the secondary target lies---but {\tt OBJID} and {\tt RELEASE} have different meanings. {\tt OBJID}, for secondary targets, records the row that a target was listed in a file of secondary targets prepared by a proposer to pass to the \dt\ pipeline, after first partitioning by {\tt BRICKID}. {\tt RELEASE}, for secondary targets, corresponds to:

\begin{equation}
\label{eqn:release}
100[X-1]+\log_2(S)~, 
\end{equation}

\noindent where $X$ is the iteration of {\tt SV$X$} during which the target was observed (see, e.g., Table~\ref{table:bitcols}); with $[X-1]=5$ used to denote the \desi\ {\tt main} survey\footnote{New or updated sets of secondary targets assigned during the Main Survey will progressively use higher (integer values) of $X$.}. Here, $S$ is the bit-value from the {\tt scnd\_mask} (see, e.g., Table~\ref{table:bitcmds}). For example, {\tt pseudo-RELEASE = 245} would denote secondary targeting bit 45 from the {\tt SV3} campaign and {\tt pseudo-RELEASE = 534} would denote secondary targeting bit 34 from the {\tt main} observing campaign. Note that secondary targets that have not adopted the primary {\tt TARGETID} can always be identified by virtue of having  {\tt RELEASE < 1000}. {\tt TARGETID} bits other than {\tt OBJID}, {\tt BRICKID} and {\tt RELEASE} are set to zero for secondary targets.

\subsubsection{The {\tt TARGETID} for Targets of Opportunity}
\label{sec:ToO}

When merited, the \desi\ survey can occasionally be suspended to pursue rapidly designed, dedicated tiles that place fibers on Targets of Opportunity (ToOs). An example science case might be to follow-up potential host galaxies of gravitational waves or neutrino events \citep[e.g.][]{Pal21}. \desi\ targeting also facilitates an observing mode where some ToOs are added to the {\em existing pool} of science targets. As with any \desi\ target class, ToOs must have a unique {\tt TARGETID} so that they can be tracked by the end-to-end pipeline. ToOs have a {\tt TARGETID} that follows the model of Figure~\ref{fig:TARGETID}, and {\tt BRICKID} has the same meaning as for general primary and secondary targets. The list of all \desi\ ToOs is maintained in a single ledger, to which new targets can be added but not removed\footnote{See E.\ Schlafly et al.\ 2023 (in preparation) for a full  description of the ToO ledger.}, and the {\tt OBJID} for ToOs is simply the row number of the corresponding target in that ledger. For ToOs, {\tt RELEASE} is always set to 9999, which is distinct from any other {\tt RELEASE} value used to encode a {\tt TARGETID}. The other possible {\tt TARGETID} bits are set to zero for ToOs.

\subsubsection{Tertiary Targets}
\label{sec:tertiary}

Occasionally, special tiles need to be rapidly scheduled for a scientific purpose that doesn't comport with the \dt\ scheme for Targets of Opportunity. To facilitate prompt observation of such tiles, \dt\ reserves {\tt RELEASE} bits equivalent to 8888 to denote targets that were ``directly'' assigned fibers. Targets that have a {\tt TARGETID} with a {\tt RELEASE} of 8888 were not processed by the \dt\ pipeline and are instead maintained and handled by the DESI {\tt fiberassign} code (A.\ Raichoor et al.\ 2023, in preparation). Such programs are sometimes referred to as ``tertiary'' programs.

\subsection{Encoding and Decoding {\tt TARGETID}}
\label{sec:decode}

The \dt\ codebase includes utility functions for extracting the bits depicted in Figure~\ref{fig:TARGETID} from a {\tt TARGETID} and for constructing a {\tt TARGETID} from its constituent bits. These functions will work for all {\em positively} valued {\tt TARGETID} integers. 

\subsubsection{\dt\ Functionality}

The specific functions in \dt\ for encoding and decoding {\tt TARGETID} can be called as follows: \\

\small{
{\tt from desitarget.targets import decode\_targetid, encode\_targetid} \\
\indent {\tt objid, brickid, release, mock, sky, gaiadr = decode\_targetid(targetid)} \\ 
\indent {\tt targetid = encode\_targetid(objid, brickid, release, mock, sky, gaiadr)} . \\ 
}

\noindent Both of the {\tt encode\_targetid} and {\tt decode\_targetid} functions are written with a high degree of flexibility. For example, {\tt targetid} can be passed to {\tt decode\_targetid} as a single integer or as an array. Similarly, {\tt encode\_targetid} can be passed a mix of arrays and integers. Additionally, none of the inputs {\em must} be passed. This makes it straightforward to encode a {\tt TARGETID} that is composed of, say, arrays of {\tt BRICKID}s and {\tt OBJID}s but only a single {\tt RELEASE} number.

\subsubsection{Negative {\tt TARGETID}s}
\label{sec:negativetid}

Some {\tt TARGETID} integers in \desi\ files have negative values. Typically, these  correspond to stuck fibers that could not be used by \desi\ to observe a science target, so were assigned to sky locations on-the-fly (see, e.g., E.\ Schlafly et al.\ 2023, in preparation). Assigning stuck fibers to sky locations frees up fibers to be assigned to science targets elsewhere in the focal plane.
The least-significant 29 bits encoded in a negative {\tt TARGETID} correspond to Declination, partitioning in equal bins in a northerly direction with the Dec = -90$^\circ$ bin corresponding to bit 0. The next-most-significant 30 bits correspond to RA, partitioned in equal bins with the RA = 0$^\circ$ bin corresponding to the least significant bit. The remaining four bits allow flexibility to record up to 16 different groups of distinct negative {\tt TARGETID}s.

The resulting schema allows unique {\tt TARGETID}s to be assigned by location at a resolution of $\sim$1.2~milliarcseconds. Negative {\tt TARGETID}s can be created and decoded using \dt\ routines as follows: \\

\small{
{\tt from desitarget.targets import decode\_negative\_targetid, encode\_negative\_targetid} \\
\indent {\tt ra, dec, group = decode\_negative\_targetid(targetid)} \\ 
\indent {\tt targetid = encode\_negative\_targetid(ra, dec, group=group)} . \\ 
}

Here, {\tt decode\_negative\_targetid(targetid)} will return the lowest-valued bin edge in RA and Dec (i.e. the ``left-hand" and ``bottom" edges of the bin). A full description of how negative {\tt TARGETID}s are handled by the \desi\ fiber assignment code is provided in A.\ Raichoor et al.\ (2023, in preparation).

\section{\desi\ Target Classes}
\label{sec:primary}

In this section, we detail each of the main \desi\ target classes and the location and structure of the files in which they are stored. We will frequently refer to columns in these files being populated by certain ``target selection bits," which are detailed in \S\ref{sec:bitmasks}.

\begin{deluxetable*}{lccr}[t]
\tablecaption{Publications that describe specific \desi\ science target classes}\label{table:tspapers}
\tablewidth{0pt}
\tablehead{
\colhead{Target selection class} & 
\colhead{Main Survey bit names\tablenotemark{a}} & 
\colhead{Brief description} &
\colhead{Selection detailed in}
}
\startdata
\sidehead{\em Dark-time targets}
 ~~~Luminous Red Galaxies & {\tt LRG} & LRG target & \citet{lrg} \\
 ~~~Emission Line Galaxies & {\tt ELG} & ELG target & \citet{elg} \\
                        & {\tt ELG\_LOP} & ELG at standard priority & \\
                        & {\tt ELG\_HIP} & ELG observed at the & \\
                        &                & priority of an LRG\tablenotemark{b} & \\
                        & {\tt ELG\_VLO} & Very-low-priority ``filler" ELG & \\
 ~~~Quasars & {\tt QSO} & Quasar target & \citet{qso} \\
\sidehead{\em Bright-time targets}
 ~~~Bright Galaxy Survey & {\tt BGS\_ANY} & Any BGS bit is set & \citet{bgs} \\
                      & {\tt BGS\_FAINT\tablenotemark{c}} & Faint BGS target & \\
                      & {\tt BGS\_BRIGHT} & Bright BGS target & \\
                      & {\tt BGS\_WISE} & AGN-like BGS target & \\              
                      & {\tt BGS\_FAINT\_HIP} & Faint BGS target prioritized & \\
                      &        & like a bright BGS target\tablenotemark{b} & \\
~~~Milky Way Survey & {\tt MWS\_ANY} & Any MWS bit is set & \citet{mws}\\
                  & {\tt MWS\_BROAD}\tablenotemark{d} & Magnitude-limited bulk sample &  \\
                  & {\tt MWS\_WD} & White dwarf &  \\
                  & {\tt MWS\_NEARBY} & Volume-limited $\sim$100\,pc sample &  \\
                  & {\tt MWS\_BHB} & Blue Horizontal Branch target &  \\
                  & {\tt MWS\_MAIN\_BLUE} & Magnitude-limited blue sample &  \\
                  & {\tt MWS\_MAIN\_RED} & Magnitude-limited red sample &  \\
\sidehead{\em Backup targets}
 ~~~~~~~~Part of the & {\tt BACKUP\_GIANT\_LOP}\tablenotemark{e}    & Candidate Giant Star & MWS et al.\ 2023 (in preparation) \\
  ~~~Milky Way Survey &                                             & observed at lower priority  &  \\
     & {\tt BACKUP\_GIANT}        & Candidate Giant Star &  \\
                          & {\tt BACKUP\_BRIGHT}   & Brighter backup target &  \\                    
                          & {\tt BACKUP\_FAINT}        & Fainter backup target &  \\
                          & {\tt BACKUP\_VERY\_FAINT}  & Even fainter backup target &  \\
\enddata
\tablenotetext{a}{Stored as bit-values in the {\tt DESI\_TARGET}, {\tt BGS\_TARGET} and {\tt MWS\_TARGET} columns. Bit-names can be converted to bit-values using the {\tt desi\_mask}, {\tt bgs\_mask} and {\tt mws\_mask} bitmasks (see \S\ref{sec:bitcols}).}
\tablenotetext{b}{Some targets with low observational priority are observed at higher priority to help characterize the survey selection function.}
\tablenotetext{c}{BGS bits other than {\tt BGS\_ANY} are stored in the {\tt BGS\_TARGET} column and {\tt bgs\_mask} bitmask.}
\tablenotetext{d}{MWS bits other than {\tt MWS\_ANY} are stored in the {\tt MWS\_TARGET} column and {\tt mws\_mask} bitmask.}
\tablenotetext{e}{BACKUP bits are stored in the {\tt MWS\_TARGET} column and {\tt mws\_mask} bitmask.}
\end{deluxetable*}
\begin{deluxetable*}{ccc}[t]
\tablecaption{Columns added to target files by the \dt\ pipeline}\label{table:tscolumns}
\tablewidth{0pt}
\tablehead{
\colhead{Column Name} & 
\colhead{Python Format} &
\colhead{Description}
}
\startdata
{\tt BRICK\_OBJID }   & {\tt i4 } & {\tt OBJID} from the Legacy Surveys sweep catalogs. \\
{\tt MORPHTYPE }     & {\tt U4 } & {\tt TYPE} from the Legacy Surveys sweep catalogs. \\
{\tt PHOTSYS }       & {\tt U1 } & `N' for {\em MzLS}/{\em BASS} photometric system, `S' for {\em DECaLS}.\tablenotemark{a} \\
{\tt TARGETID }      & {\tt i8 } &  See \S\ref{sec:targetid}. \\
{\tt DESI\_TARGET }   & {\tt i8 } & Column name differs for SV; see \S\ref{sec:bitcols}. \\
{\tt BGS\_TARGET }    & {\tt i8 } & Column name differs for SV; see \S\ref{sec:bitcols}. \\
{\tt MWS\_TARGET }    & {\tt i8 } & Column name differs for SV; see \S\ref{sec:bitcols}. \\
{\tt SCND\_TARGET }   & {\tt i8 } & Column name differs for SV; see \S\ref{sec:bitcols}. \\
{\tt SUBPRIORITY }   & {\tt f8 } &  Random number (0--1) used to break fiber-assignment ties.\\
{\tt OBSCONDITIONS } & {\tt i8 } & See \S\ref{sec:obscon}. \\
{\tt PRIORITY\_INIT } & {\tt i8 } & See E.\ Schlafly et al.\ 2023 (in preparation). \\
{\tt NUMOBS\_INIT }   & {\tt i8 } & See E.\ Schlafly et al.\ 2023 (in preparation). \\
{\tt HPXPIXEL }      & {\tt i8 } & The {\tt nside = 64} HEALPixel of the target.\tablenotemark{b} \\
\enddata
\tablenotetext{a}{{\em BASS}, {\em MzLS} and {\em DECaLS} are the individual surveys that comprise the \desi\ Legacy Imaging Surveys \citep[e.g.][]{Dey19}. The photometric system ({\tt bok} for {\em BASS}, {\tt mosaic} for {\em MzLS} and {\tt decam} for {\em DECaLS}) can be derived from the {\tt RELEASE} number according to \url{https://www.legacysurvey.org/release/}.}
\tablenotetext{b}{This {\tt nside} is stored in the file header as {\tt HPXNSIDE}.}
\end{deluxetable*}

\subsection{Primary Science Targets}

The selection and optimization of \desi\ science targets for SV3 and the Main Survey are described in a series of papers, listed in Table~\ref{table:tspapers}. 

\subsubsection{Bright-time and Dark-time Targets}
\label{sec:btanddt}

Bright-time and dark-time science targets are publicly available online for SV1\footnote{\urlstub{dr9/0.51.0/targets/sv1/resolve/}}, SV2\footnote{\urlstub{dr9/0.53.0/targets/sv2/resolve/}}, SV3\footnote{\urlstub{dr9/0.57.0/targets/sv3/resolve/}} and the Main Survey\footnote{\urlstub{dr9/1.1.1/targets/main/resolve/}}. Each url links to {\tt bright} and {\tt dark} sub-directories which contain bright-time (MWS, BGS) and dark-time (ELG, LRG, QSO) targets, respectively. Note that operational procedures were in flux during SV1, meaning that a (very small) subset of SV1 targets can only be found in earlier versions of the target files than {\tt 0.51.0}. Similarly, a few hundred targets from the DESI Main Survey are only available in the {\tt 1.0.0} version of the target files (see \S\ref{sec:targrep}).

The {\tt bright}/{\tt dark} sub-directories host files containing targets that are only selected by bright-time/dark-time target selection algorithms. But, every target in each sub-directory has {\em all of the appropriate targeting bits} populated. Consider, for instance, a QSO target that is {\em also} targeted by the MWS survey. Such a target will appear in the appropriate file in {\em both} of the {\tt bright}/{\tt dark} sub-directories with the same bits set in the {\tt DESI\_TARGET} and {\tt MWS\_TARGET} columns. However, a QSO target that is {\em not} also targeted by a bright-time survey will {\em only} appear in the appropriate file in the {\tt dark} sub-directory. Similarly, a bright-time target that is not also targeted as part of the dark-time survey will only appear in the appropriate file in the {\tt bright} sub-directory.

Files in the {\tt bright}/{\tt dark} sub-directories have names of the form {\tt AAA}targets-{\tt OBSCON}-hp-{\tt HPX}.fits. Here, {\tt AAA} is {\tt sv1}, {\tt sv2} or {\tt sv3} for SV files and is omitted for Main Survey files, {\tt OBSCON} is either {\tt dark} or {\tt bright}, and {\tt HPX} refers to the {\tt nside = 8} integer of the HEALPixel that contains the targets.\footnote{This {\tt nside} is stored in the file header as {\tt FILENSID}.}

The contents of each file described in this section are similar. Information is stored in FITS format \citep{FITS} and data is only included in extension 1 of the FITS file. Most of the data columns are directly derived from the DR9 ``sweep" and ``light curve sweep" catalogs\footnote{\lsurl{dr9/files/\#sweep-catalogs-region-sweep}.} from which \desi\ Main Survey targets were selected. Additional columns that are derived by \dt\ itself are detailed in Table~\ref{table:tscolumns}.

\begin{deluxetable*}{crrrrrrr}[t]
\tablecaption{Densities (deg$^{-2}$) of principal Main Survey target classes in dark time}\label{table:targdensdark}
\tablewidth{0pt}
\tablehead{
\colhead{Bit name} & 
\colhead{\tt BGS\_ANY} & 
\colhead{\tt ELG} &
\colhead{\tt LRG} &
\colhead{\tt MWS\_ANY} &
\colhead{\tt QSO} &
\colhead{\tt STD\_BRIGHT\tablenotemark{a}} &
\colhead{\tt STD\_FAINT\tablenotemark{a}}
}
\startdata
{\tt BGS\_ANY}      & 79.8\tablenotemark{b} &    1.0\tablenotemark{c} &  78.4 &   0.5 &   0.9 &  0.0 &   0.0 \\
{\tt ELG}           &      & 2394.3 &   2.3 &   0.0 & 108.5 &  0.0 &   0.0 \\
{\tt LRG}           &      &        & 625.7 &   0.4 &   4.0 &  0.0 &   0.0 \\
{\tt MWS\_ANY}      &      &        &       & 119.0 &   7.6 & 51.9 & 104.2 \\
{\tt QSO}           &      &        &       &       & 319.8 &  0.0 &   0.0 \\
{\tt STD\_BRIGHT}   &      &        &       &       &       & 54.9 &  54.9 \\
{\tt STD\_FAINT}    &      &        &       &       &       &      & 107.2 \\
\enddata
\tablenotetext{a}{{\tt STD\_BRIGHT} and {\tt STD\_FAINT} are the main standard star classes intended for use in bright time and dark time, respectively (see \S{\ref{sec:standards}}).}
\tablenotetext{b}{On-diagonal elements represent the density for a target class, with areas derived using the {\tt desitarget} weight maps (see \S\ref{sec:pixweight}).}
\tablenotetext{c}{Off-diagonal elements represent the overlap in density between two target classes.}
\end{deluxetable*}
\begin{deluxetable*}{crrrrrrr}[t]
\tablecaption{Densities (deg$^{-2}$) of principal Main Survey target classes in bright time}\label{table:targdensbright}
\tablewidth{0pt}
\tablehead{
\colhead{Bit name} & 
\colhead{\tt BGS\_ANY} & 
\colhead{\tt ELG} &
\colhead{\tt LRG} &
\colhead{\tt MWS\_ANY} &
\colhead{\tt QSO} &
\colhead{\tt STD\_BRIGHT\tablenotemark{a}} &
\colhead{\tt STD\_FAINT\tablenotemark{a}}
}
\startdata
{\tt BGS\_ANY}      & 1389.7\tablenotemark{b} & 1.0\tablenotemark{c} &  78.4 &   0.6 &   0.9 &  0.0 &   0.0 \\
{\tt ELG}           &      &    1.0 &   0.4 &   0.0  &   0.0 &  0.0 &   0.0 \\
{\tt LRG}           &      &        &  78.8 &   0.4  &   0.3 &  0.0 &   0.0 \\
{\tt MWS\_ANY}      &      &        &       & 1910.5 &   7.6 & 51.9 & 104.2 \\
{\tt QSO}           &      &        &       &        &   8.0 &  0.0 &   0.0 \\
{\tt STD\_BRIGHT}   &      &        &       &        &       & 54.9 &  54.9 \\
{\tt STD\_FAINT}    &      &        &       &        &       &      & 107.2 \\
\enddata
\tablenotetext{a}{{\tt STD\_BRIGHT} ({\tt STD\_FAINT}) is the principal standard star class (see \S{\ref{sec:standards}}) intended for use in bright time (dark time).}
\tablenotetext{b}{On-diagonal elements represent the density for a target class, with areas derived using the {\tt desitarget} weight maps (see \S\ref{sec:pixweight}).}
\tablenotetext{c}{Off-diagonal elements represent the overlap in density between two target classes.}
\end{deluxetable*}

In Table~\ref{table:targdensdark} and Table~\ref{table:targdensbright} we show the densities of the principal \desi\ Main Survey bright-time and dark-time target classes. We also show how the densities of these target classes overlap (i.e.\ whether a target is included in {\em both} of two target classes). For example, Table~\ref{table:targdensbright} demonstrates that $\sim$0.4\% (7.6/1910.5) of MWS targets are also quasar targets. This overlap is driven by the {\tt MWS\_MAIN\_BLUE} sample described in \S4.1 of \citet{mws}, which contains blue point sources that resemble quasars in imaging. Densities in these tables are derived from the target files discussed in this section, with areal weights derived using the {\tt desitarget} pixelized weight maps detailed in \S\ref{sec:pixweight}. The resulting matrices of densities can also be seen on the front page of the Quality Assurance (QA) web pages described in \S\ref{sec:QA}.

\subsubsection{Backup Targets}
\label{sec:backup}

In addition to other campaigns, \desi\ pursues a program of ``backup" targets when observing conditions are of insufficient quality to collect signal on bright-time or dark-time science targets (see also \S\ref{sec:obscon}). The principal backup targets are listed in the final part of Table~\ref{table:tspapers} and will be detailed in a forthcoming paper (see MWS et al.\ 2023, in preparation). Backup targets differ from other science targets in that they are selected solely using {\em Gaia} information rather than imaging from the Legacy Surveys. As such, the directory structure for the files that contain backup targets differs slightly from that of other science targets, having {\tt gaiadr2} in the directory construction instead of {\tt dr9}\footnote{For instance, SV3 backup targets are at \urlstub{gaiadr2/0.57.0/targets/sv3/resolve/}.}. For versions of \dt\ starting with {\tt 0.52.0}, backup targets are stored in a {\tt backup} sub-directory but prior to {\tt 0.52.0}, this sub-directory was instead called {\tt supp} (for ``supplemental"). Backup targets used for the \desi\ Main Survey are derived from a slightly later release of the \dt\ code than for other target classes.\footnote{\urlstub{gaiadr2/2.2.0/targets/main/resolve/}.}

\subsubsection{Overlapping Imaging: The North/South ``Resolve''}
\label{sec:resolve}

The majority of \desi\ targets are selected from the Legacy Surveys, which comprises  three individual imaging campaigns called {\em BASS, MzLS and DECaLS}. Because these surveys overlap \citep[see, e.g., Figure 1 of][]{Dey19}, it is crucial to decide which imaging to use in each area of the sky. To help decide this, the \dt\ pipeline adds a {\tt PHOTSYS} column to targeting files (see Table~\ref{table:tscolumns}), which is set to `N' for targets derived using imaging from the ``northern'' Legacy Surveys ({\em BASS} and {\em MzLS}) and `S' for targets derived using DECaLS imaging. A function called {\tt targets.resolve} in the \dt\ pipeline then uses {\tt PHOTSYS} to determine (or ``resolve'') which imaging to adopt in a given area of the \desi\ footprint.

The \dt\ resolve function first defines the ``northern" sky to comprise areas that are {\em both} north of the Galactic plane (in Galactic coordinates) and north of $\mathrm{Dec} \geq 32.375^\circ$ (in equatorial coordinates). Any area that does not meet both of these criteria is considered to be ``southern" area \citep[again, Figure 1 of][may be helpful to visualize these areas]{Dey19}. The \dt\ resolve function then ensures unique imaging by only retaining targets that are in the northern sky area and that have {\tt PHOTSYS == `N'} set, or that are in the southern sky area and have {\tt PHOTSYS == `S'} set. Note that no resolve is necessary for targets---such as backup targets---that are selected purely from {\em Gaia} data.

The urls from which to retrieve targets that are provided in \S\ref{sec:btanddt} all terminate with a {\tt resolve} directory, indicating that they have been resolved by \dt\ into unique targets in areas where the individual Legacy Surveys overlap. It is possible to use the \dt\ pipeline to generate {\em all} targets in overlapping imaging areas rather than resolving them, and some (very early) versions of \desi\ targeting files were processed with this option. Such targets end up in a directory structure that includes {\tt noresolve} instead of {\tt resolve}.

\begin{deluxetable}{c}[t]
\tablecaption{Columns that use {\em Gaia} EDR3 instead of DR2 when {\tt GAIASUB} is set in a file header. Columns are as for the Legacy Surveys sweep files.}\label{table:edr3}
\tablewidth{0pt}
\tablehead{
\colhead{Column name}
}
\startdata
{\tt PARALLAX} \\
{\tt PARALLAX\_IVAR} \\
{\tt PMRA} \\
{\tt PMRA\_IVAR} \\
{\tt PMDEC} \\
{\tt PMDEC\_IVAR} \\
{\tt GAIA\_DUPLICATED\_SOURCE} \\
{\tt GAIA\_ASTROMETRIC\_PARAMS\_SOLVED} \\
{\tt GAIA\_ASTROMETRIC\_SIGMA5D\_MAX} \\
{\tt GAIA\_ASTROMETRIC\_EXCESS\_NOISE} \\
\enddata

\end{deluxetable}

\subsubsection{Gaia DR2 or Gaia EDR3?}
\label{sec:gaiasub}

Version {\tt 0.58.0} of the \dt\ code introduced the option of partially using {\em Gaia} Early Data Release 3 \citep[EDR3;][]{GaiaEDR3} for target selection, with a view to optimizing astrometry-based selection of MWS targets without changing {\em Gaia} imaging information for selections used by other target classes. Target files processed with this option have a {\tt GAIASUB} value in the FITS header that is set to {\tt True}. EDR3 values are obtained by coordinate-matching the Legacy Surveys sweep catalogs used for target selection (see \S\ref{sec:btanddt}) and EDR3 at 0.2\arcsec, after first accounting for EDR3 proper motions. The EDR3 columns that are substituted for {\em Gaia} in such target files are listed in Table~\ref{table:edr3}. 

So, broadly, SV targets were selected using {\em Gaia} DR2 and Main Survey targets were selected using imaging quantities from {\em Gaia} DR2 and astrometric quantities from {\em Gaia} EDR3. An important caveat to this statement is that, as of the time of writing---corresponding to version {\tt 2.5.0} of \dt---{\em backup targets (\S\ref{sec:backup}) still always use {\em Gaia} DR2 for all quantities}. This is due to a bug where the {\em Gaia} EDR3 quantities are {\em not} substituted for backup targets  even though the {\tt GAIASUB} keyword {\em is} set to {\tt True}.

\begin{deluxetable*}{cc}[t]
\tablecaption{Target Selection for the {\tt STD\_BRIGHT} and {\tt STD\_FAINT}  target classes}\label{table:stdbright}
\tablewidth{0pt}
\tablehead{
\colhead{Reason for Selection} & 
\colhead{How Selection is Applied}
}
\startdata
Not masked &	{\tt MASKBITS}\tablenotemark{a,b} is not {\tt BRIGHT} or {\tt GALAXY} \\
Point-like & {\tt TYPE == "PSF"} \\
Isolated &	${\tt FRACFLUX\_X}\tablenotemark{c} < 0.01$ \\
Measured flux & ${\tt FLUX\_IVAR\_X} > 0$ \\
Observed &	${\tt NOBS\_X} > 0$ \\
Low fraction of masked pixels &	${\tt FRACMASKED\_X} < 0.6$ \\
Color of halo turnoff or bluer\tablenotemark{d}  &	$(r-z)<0.2$ AND $0.0 < (g-r) < 0.35$ \\
Reasonably bright in {\em Gaia}\tablenotemark{e} & $16 \leq {\tt PHOT\_G\_MEAN\_MAG} < 18$ \\
$Bp-Rp$ color is measured & ${\tt GAIA\_PHOT\_BP\_MEAN\_MAG} - {\tt GAIA\_PHOT\_RP\_MEAN\_MAG}$ is not {\tt NaN} \\
No astrometry issues &	${\tt GAIA\_ASTROMETRIC\_EXCESS\_NOISE} < 1$ AND \\
                     & ${\tt GAIA\_ASTROMETRIC\_PARAMS\_SOLVED} == 31$ \\
Proper motions are finite &	Neither {\tt PMRA} or {\tt PMDEC} is {\tt NaN} \\
Parallax smaller than 1\,mas & ${\tt PARALLAX} < 1$ \\
Proper motion larger than 2\,mas\,yr$^{-1}$ & $ \sqrt{{\tt PMRA}^2 + {\tt PMDEC}^2} > 2$ \\
Unique source in {\em Gaia} &	${\tt GAIA\_DUPLICATED\_SOURCE} == 0$ \\
\enddata
\tablenotetext{a}{Parameters that appear in this table are derived from the Legacy Surveys sweep files at \url{https://www.legacysurvey.org/dr9/files/\#sweep-catalogs-region-sweep}.}
\tablenotetext{b}{The Legacy Surveys {\tt MASKBITS} mask is described at \url{https://www.legacysurvey.org/dr9/bitmasks/}.}
\tablenotetext{c}{In this table, {\tt X} denotes that the criterion is applied for {\em all} bands from the Legacy Surveys ($g$, $r$ and $z$).}
\tablenotetext{d}{$g$, $r$, $z$ denote {\tt FLUX\_G}, {\tt FLUX\_R}, {\tt FLUX\_Z} converted to magnitudes and corrected for Galactic extinction.} 
\tablenotetext{e}{The only difference between the {\tt STD\_BRIGHT} and {\tt STD\_FAINT} target classes is that {\tt STD\_FAINT} extends across $16 \leq {\tt PHOT\_G\_MEAN\_MAG} < 19$.}
\end{deluxetable*}
\begin{deluxetable*}{cc}[t]
\tablecaption{Target Selection for the {\tt GAIA\_STD\_BRIGHT} and {\tt GAIA\_STD\_FAINT} target classes}\label{table:gaiastdbright}
\tablewidth{0pt}
\tablehead{
\colhead{Reason for Selection} & 
\colhead{How Selection is Applied}
}
\startdata
Point-like & ${\tt GAIA\_ASTROMETRIC\_EXCESS\_NOISE} < 10^{0.5}$  \\
Isolated &	No other {\em Gaia} source within 3.5\arcsec \\
Color of halo turnoff or bluer\tablenotemark{a} & $0.2 < (Bp - Rp) < 0.9$ AND \\
           & $(G - Bp) > -0.5(Bp - Rp)$ AND \\
           & $(G - Bp) < 0.3-0.5(Bp - Rp)$  \\
Reasonably bright in {\em Gaia}\tablenotemark{b} & $16 \leq {\tt PHOT\_G\_MEAN\_MAG} < 18$ \\
$Bp-Rp$ color is measured & ${\tt GAIA\_PHOT\_BP\_MEAN\_MAG} - {\tt GAIA\_PHOT\_RP\_MEAN\_MAG}$ is not {\tt NaN} \\
No astrometry issues &	${\tt GAIA\_ASTROMETRIC\_EXCESS\_NOISE} < 1$ AND \\
                     & ${\tt GAIA\_ASTROMETRIC\_PARAMS\_SOLVED} == 31$ \\
Proper motions are finite &	Neither {\tt PMRA} or {\tt PMDEC} is {\tt NaN} \\
Parallax smaller than 1\,mas & ${\tt PARALLAX} < 1$ \\
Proper motion larger than 2\,mas\,yr$^{-1}$ & $ \sqrt{{\tt PMRA}^2 + {\tt PMDEC}^2} > 2$ \\
Unique source in {\em Gaia} &	${\tt GAIA\_DUPLICATED\_SOURCE} == 0$ \\
\enddata
\tablenotetext{a}{In this table, $Bp$, $G$, $Rp$ denote versions of {\tt GAIA\_PHOT\_BP\_MEAN\_MAG}, {\tt GAIA\_PHOT\_G\_MEAN\_MAG}, {\tt GAIA\_PHOT\_RP\_MEAN\_MAG} that have been corrected for Galactic extinction using the prescription in Table 1 of \citet{Bab18}.}
\tablenotetext{b}{The only difference between the {\tt GAIA\_STD\_BRIGHT} and {\tt GAIA\_STD\_FAINT} target classes is that {\tt GAIA\_STD\_FAINT} extends across $16 \leq {\tt PHOT\_G\_MEAN\_MAG} < 19$.}
\end{deluxetable*}

\subsection{Standard Stars}
\label{sec:standards}

The \desi\ survey requires flux standards for spectrophotometric calibration. To achieve this goal, \dt\ defines three principal standard star classes to be used inside of the footprint of the Legacy Surveys imaging, with bit-names {\tt STD\_BRIGHT}, {\tt STD\_FAINT} (see also Table~\ref{table:targdensdark} and Table~\ref{table:targdensbright}) and {\tt STD\_WD}\footnote{``{\tt WD}'' denotes a white dwarf.}, and three classes for use outside of the Legacy Surveys imaging footprint, {\tt GAIA\_STD\_BRIGHT}, {\tt GAIA\_STD\_FAINT} and {\tt GAIA\_STD\_WD}. The {\em Gaia} standard star selections are designed to approximate the principal standard star classes but using imaging from {\em Gaia} in place of Legacy Surveys quantities. The {\tt STD\_WD} and {\tt GAIA\_STD\_WD} classes are identical to each other---as well as to the {\tt MWS\_WD} class described in \citet{mws}---so won't be further detailed here.

The {\tt STD\_BRIGHT} and {\tt STD\_FAINT} target classes are designed to select Main Sequence F-stars in a similar fashion to the BOSS selection of spectrophotometric standards\footnote{\url{https://www.sdss.org/dr12/algorithms/boss_std_ts/}}. Color constraints differ somewhat from those applied in BOSS, though, both because DR9 of the Legacy Surveys does not include deep $u$- and $i$-band imaging (unlike the SDSS) and to ensure sufficient flux standards across the \desi\ focal plane. The overall approach introduces a small fraction of A-type stars, while retaining sufficiently precise flux calibration to measure signal in the Lyman-$\alpha$ Forest. Broadly, the {\tt STD\_BRIGHT} target class is intended to be used for calibration during bright-time programs and the {\tt STD\_FAINT} target class is intended for use during dark time.

The {\tt STD\_BRIGHT} flux standards are selected as detailed in Table~\ref{table:stdbright}. The only difference between {\tt STD\_BRIGHT} and {\tt STD\_FAINT} targets is that {\tt STD\_FAINT} targets extend over the magnitude range $16 \leq G < 19$ instead of $16 \leq G < 18$ (in {\em Gaia} $G$-band)\footnote{Standards appropriate for bright-time were also intended to be used in dark-time, but not vice-versa, meaning that the {\tt STD\_FAINT} class is a ``fainter'' extension of the {\tt STD\_BRIGHT} selection. Hence the name ``{\tt STD\_FAINT}'' rather than, say, ``{\tt STD\_DARK}.''}. The selection of the {\tt GAIA\_STD\_BRIGHT} and {\tt GAIA\_STD\_FAINT} flux standards is detailed in Table~\ref{table:gaiastdbright}. Again, {\tt GAIA\_STD\_BRIGHT} and {\tt GAIA\_STD\_FAINT} targets only differ because {\tt GAIA\_STD\_FAINT} standards extend over the magnitude range $16 \leq G < 19$. The proper motion cuts listed in Tables~\ref{table:stdbright} and \ref{table:gaiastdbright} are included to help avoid stars in the thick disk of the Milky Way in favor of a more metal-poor halo turn-off population. Thick disk stars are more numerous at distances of a few kiloparsecs from the Sun---but the halo is hotter. Halo stars therefore typically have larger proper motions than thick disk stars at intermediate distances.

The algorithms used to select standard stars remained the same throughout \desi\ SV and the Main Survey. {\em Gaia}-only standards ({\tt GAIA\_STD\_BRIGHT}, etc.) weren't introduced until version {\tt 0.48.0} of \dt---and weren't actually finalized until version {\tt 0.50.0}---meaning that these standards do not appear in target files that {\em pre-date} SV1. As with other target classes (see \S\ref{sec:gaiasub}) the option to substitute {\em Gaia} EDR3 astrometric data became available in version {\tt 0.58.0} of the \dt\ code, and this option was used to produce targeting files for the \desi\ Main Survey.

Because the {\em Gaia}-only standards were designed for observations outside of the Legacy Surveys imaging footprint, the data model for these targets is slightly different to that of the principal flux standards. First, the {\em Gaia}-only bits ({\tt GAIA\_STD\_BRIGHT}, etc.) can be accessed via the {\tt mws\_mask} bitmask and are stored in the
{\tt MWS\_TARGET} column (see Tables~\ref{table:bitcmds} and \ref{table:bitcols}), whereas the principal standards ({\tt STD\_BRIGHT}, etc.) can be accessed via the {\tt desi\_mask} bitmask and are stored in the {\tt DESI\_TARGET} column. Second, the {\em Gaia}-only standards, as is the case for all {\em Gaia}-only targets, reside in the backup files outlined in \S\ref{sec:backup} rather than the bright-time and dark-time targeting files outlined in \S\ref{sec:btanddt}.

\subsection{Guiding, Focusing and Alignment (GFA) targets}
\label{sec:gfas}

The \desi\ instrument requires samples of stars to use for guiding, focusing and alignment. Collectively, we will refer to these as GFA targets, or just ``GFAs.'' The \dt\ pipeline includes a module called {\tt gfa} specifically for assembling these targets. GFAs are publicly available online in directories accompanying those described in \S\ref{sec:btanddt}, where the directory names terminate with {\tt gfas} instead of {\tt targets}\footnote{For example, for the Main Survey the appropriate url is \urlstub{dr9/1.1.1/gfas/}}. Within a {\tt gfas} directory, files have names of the form gfas-hp-{\tt HPX}.fits. Here, {\tt HPX} refers to the {\tt nside = 8} integer of the HEALPixel that contains the GFAs.\footnote{As for science targets, this {\tt nside} is stored in the file header as {\tt FILENSID}.}

The selection of GFAs did not significantly change between iterations of SV and the \desi\ Main Survey. GFA targets essentially comprise all sources in {\em Gaia} limited to $G < 21$ in the {\em Gaia} $G$-band. We include extra imaging information for {\em Gaia} sources that are also in the Legacy Surveys sweep catalogs\footnote{\lsurl{dr9/files/\#sweep-catalogs-region-sweep}.} (i.e.\ that have {\tt REF\_ID > 0}) and that are {\em not} part of the 2020 Siena Galaxy Atlas (J.\ Moustakas et al.\ 2023, in preparation). This ensures that high-precision fluxes and morphologies are retained for GFAs in the primary \desi\ targeting footprint to make quality cuts downstream of \dt, if needed.

Between the SV and Main Survey phases of \desi\ the {\tt desitarget.gfa} code was updated to include the option of substituting {\em Gaia}-based values from {\em Gaia} EDR3 in place of values from {\em Gaia} DR2 (see also \S\ref{sec:gaiasub}). GFA targets for the \desi\ Main Survey included this substitution. In general, files of GFAs that were run with {\em Gaia} EDR3 have a {\tt GAIADR} value in the FITS header that is set to ``{\tt edr3}''. For GFAs that have a match in the Legacy Surveys sweep catalogs, only the columns in Table~\ref{table:edr3} are ever updated from {\em Gaia} DR2 to {\em Gaia} EDR3. For GFAs that do not match a Legacy Surveys source (i.e.\ typically objects outside of the Legacy Surveys imaging footprint) {\em all} {\em Gaia}-derived quantities are updated to EDR3.

Some sources that appear in {\em Gaia}---particularly in DR2---do not have measured proper motions. As astrometric information is essential for \desi\ guiding, the GFAs incorporate proper motions from the First U.S. Naval Observatory Robotic Astrometric Telescope Catalog \citep[URAT;][]{URAT} where {\em Gaia} proper motions are missing. Only proper motions for URAT sources that match a {\em Gaia} source within 0.5\arcsec\ are substituted for missing {\em Gaia} values. The \dt\ code has the option of turning off this URAT-substitution mechanism. GFA targets that {\em did} include substitution of URAT proper motions in place of missing {\em Gaia} values can be identified by having {\tt NOURAT} set to {\tt False} in the FITS header of the GFA file.

All of the columns in the \dt\ GFA files are derived from the Legacy Surveys sweep catalogs\footnote{\lsurl{dr9/files/\#sweep-catalogs-region-sweep}} or are provided in Table~\ref{table:tscolumns}, with the exception of {\tt URAT\_ID}, {\tt URAT\_SEP} and {\tt GAIA\_PHOT\_G\_N\_OBS}. The {\tt URAT\_ID} column records the recommended identifier for URAT1 sources (see \url{http://cdsarc.u-strasbg.fr/ftp/I/329/ReadMe}), {\tt URAT\_SEP} is the separation (in arcseconds) between a URAT source and a GFA target, and {\tt GAIA\_PHOT\_G\_N\_OBS} records the number of observations in $G$ band from {\em Gaia} (which is useful for characterizing variable sources downstream of \dt). 

In addition, the {\tt MORPHTYPE} column described in Table~\ref{table:tscolumns} has a different provenance for GFA targets that do not have a match in the Legacy Surveys. For such {\em Gaia}-only GFAs, a source is recorded as point-like if 

\begin{equation}
\label{eqn:gaiamorph}
\begin{split}
[G \leq 19.&~\mathrm{AND}~AEN < 10^{A}]~\mathrm{OR} \\ 
[G \geq 19.&~\mathrm{AND}~AEN < 10^{A + 0.2(G - 19.)}]
\end{split}
\end{equation}

\noindent where $G$ denotes {\em Gaia} $G$-band magnitude (without any correction for Galactic extinction; {\tt GAIA\_PHOT\_G\_MEAN\_MAG}), $AEN$ denotes {\tt GAIA\_ASTROMETRIC\_EXCESS\_NOISE}, and $A$ is 0.5 for {\em Gaia} DR2 or 0.3 for EDR3. Sources which do (do not) meet the criterion in Eqn.\,\ref{eqn:gaiamorph} are designated {\tt GPSF} ({\tt GGAL}) in the {\tt MORPHTYPE} column.

\subsection{Blank-sky Locations}
\label{sec:blankskies}

The \desi\ fiber allocation software (A.\ Raichoor et al.\ 2023, in preparation) assigns fibers to empty or ``blank'' locations that the \desi\ spectroscopic pipeline \citep{spec} uses to perform sky subtraction. To assign these fibers, \desi\ operations utilizes pre-calculated lists of sky locations that are expected to include minimal flux from astronomical sources when integrated over a \desi\ fiber. The \dt\ pipeline\footnote{Using a module called {\tt skyfibers}.} assembles such lists of blank sky locations using two different techniques, a pixel-level approach and a {\em Gaia}-avoidance approach, which we describe in this section. 

In addition, as of the Main Survey, \desi\ operations software determines, on-the-fly, whether stuck fibers in the \desi\ focal plane would collect significant flux during survey observations. Where possible, these stuck fibers then become part of the sky budget, which releases other fibers that can then be used to observe science targets (see \S\ref{sec:negativetid}). Reassigning sky locations on-the-fly is detailed in A.\ Raichoor et al.\ (2023, in preparation) and E.\ Schlafly et al.\ (2023, in preparation).

\subsubsection{Pixel-based Sky Locations}
\label{sec:skies}

The {\tt legacypipe} software used by the Legacy Surveys to extract sources from astronomical images operates on areas, called ``blobs,'' in which pixels have signal that exceeds a certain detection threshold. The {\tt legacypipe} code builds maps where pixels that lie in a particular blob are set to a positive integer that encodes that blob and pixels that aren't in any blob are set to -1. These ``blob maps'' are created for each brick of the Legacy Surveys and are available in the {\tt metrics} sub-directory of the DR9 Legacy Surveys release (D.\ Schlegel et al.\ 2023, in preparation), with names resembling {\tt metrics/000/blobs-0003m015.fits.gz}\footnote{For additional information regarding the Legacy Surveys maps see, e.g., \lsurl{dr9/files/\#image-stacks-region-coadd}.}. The \dt\ code adapts these blob maps by setting all pixels that do (do not) have a value of -1 to {\tt True} ({\tt False}), essentially building a pixel-map that stores values of {\tt True} where no sources are detected. The resulting source-detection-maps share the Legacy Surveys native resolution for a brick of $3600 \times 3600$ pixels, at $0.262\,\arcsec\,\mathrm{ pixel}^{-1}$. So, each map contains 12.96M pixels spread across the approximately $0.0623\,\mathrm{deg}^2$ unique area of a brick.

The \dt\ pipeline produces catalogs of pixel-level blank locations at a user-specified density of sky positions. The density used for the Main Survey, for instance, was 18{,}000 sky locations per square degree, which is easily sufficient to fill the \desi\ focal plane\footnote{This quantity is stored as {\tt NPERSDEG} in the header of files produced by \dt.}. The \dt\ code then performs a binary erosion on the source-detection-maps (in each unique brick) to achieve the required density. For example, if 18{,}000 $\mathrm{deg}^{-2}$ sky locations are desired in a $0.0623\,\mathrm{deg}^2$ brick, then a total of about 1120 locations are needed---meaning a brick needs to be eroded until it has dimensions of approximately $34 \times 34$. In each step of the erosion, in a given grid cell, \dt\ calculates the distance (in pixels) of the pixel that is farthest from a detected source; we will refer to this quantity as {\tt BLOBDIST}. At the conclusion of the erosion procedure, a blank sky target is placed in each grid cell at the location with the maximum value of {\tt BLOBDIST}. If a grid cell completely overlaps pixels that include source detections, then {\tt BLOBDIST} will be equal to 0, meaning that there are no completely blank sky locations in that grid.

The \dt\ pipeline then uses the {\tt photutils} \citep{photutils} routines {\tt CircularAperture} and {\tt aperture\_photometry} to photometer flux at each calculated sky location. Fluxes are measured in the per-filter {\tt image} and {\tt invvar} stacks provided by the Legacy Surveys\footnote{\lsurl{dr9/files/\#image-stacks-region-coadd}}. The measured fluxes, and their associated inverse variances, are then stored as quantities called {\tt FIBERFLUX\_X} and {\tt FIBERFLUX\_IVAR\_X}, where {\tt X} refers to each of the Legacy Surveys optical bands ($g$, $r$ and $z$). The measured flux quantities can have multiple values if apertures of different radii are specified when running \dt\footnote{The applied apertures are stored in a series of values called {\tt AP0}, {\tt AP1}, {\tt AP2}, etc., in the header of files produced by \dt.}, though processing for the \desi\ SV and Main Survey files exclusively adopted a $0.75\arcsec$ aperture, corresponding to the radius of a \desi\ fiber.

Given the high density of sky locations required by \desi, it is inevitable that apertures placed at certain positions will not be empty. The \dt\ pipeline therefore labels two distinct types of sky locations: particularly good positions and more dubious ones. Sky locations that have any of {\tt FIBERFLUX\_G == 0} or {\tt FIBERFLUX\_IVAR\_G == 0} or {\tt FIBERFLUX\_R == 0} or {\tt FIBERFLUX\_IVAR\_R == 0} or {\tt BLOBDIST == 0} are referred to as ``bad'' skies, and are assigned the bit-name {\tt BAD\_SKY}\footnote{All sky-related bits appear in the {\tt desi\_mask} bitmask and are stored in the {\tt DESI\_TARGET} column (see \S\ref{sec:bitmasks})}. Sky locations that do not meet these criterion are referred to just as ``skies'' and are assigned the bit-name {\tt SKY}. The logic behind the criteria that determine whether a sky location is bad is that locations with zero flux or infinite flux variance in the $g$ or $r$ bands are typically missing entirely from the Legacy Surveys---rather than genuinely having flux of exactly zero---and that sky locations with {\tt BLOBDIST == 0} certainly intersect a source, as described above.

Pixel-level sky locations are derived within Legacy Surveys bricks. It is important to note that this {\em generally contrasts with other files produced by \dt}, which are stored within HEALPixels. Therefore, before using the \dt\ pipeline to make supplemental sky locations by avoiding bright stars in {\em Gaia}---which is discussed in the next section of this manuscript---it is important to recast the derived sky locations from a brick-based scheme to a HEALPixel-based scheme. The \dt\ function that can be used to achieve this is called {\tt skyfibers.repartition\_skies}, and how this function is applied is detailed in the tutorial linked to in \S\ref{sec:tutorial}.

\subsubsection{Supplemental Sky Locations}
\label{sec:suppskies}

To account for the fact that certain areas of the sky covered by \desi\ might be outside of---or missing in---the Legacy Surveys, the \dt\ pipeline generates an additional catalog of sky positions using {\em Gaia}. These ``supplemental'' sky locations are produced at random positions across the sky within HEALPixels at a user-specified density\footnote{As with other sky locations, this quantity is stored as {\tt NPERSDEG} in the header of files produced by \dt.}. For SV and Main Survey files, this density was chosen to be 18{,}000\,${\mathrm{deg}^{-2}}$ to match the density at which pixel-based sky locations were generated. Supplemental sky locations are then removed if they are within a radius of 2\arcsec\ of any source in {\em Gaia} DR2. Finally, to limit the total density of sky locations that could be used by the \desi\ fiber assignment software, supplemental skies are removed if they share a HEALPixel with a pixel-based sky location at {\tt nside = 4096}, which corresponds to a resolution of approximately 18{,}000\,${\mathrm{deg}^{-2}}$.

Quantities generated for pixel-based sky locations are also ``mocked up'' for supplemental sky locations. The {\tt BLOBDIST} value is set to be the (2\arcsec) radius for avoiding {\em Gaia} sources, divided by the 0.262\arcsec\ pixel-scale of the Legacy Surveys imaging. All {\tt FIBERFLUX\_X} and {\tt FIBERFLUX\_IVAR\_X} quantities (see \ref{sec:skies}) are set to -1. Finally, supplemental skies are assigned the bit-name {\tt SUPP\_SKY}; as with other sky locations, this bit is set in the  {\tt DESI\_TARGET} column and can be extracted using the {\tt desi\_mask} bitmask (see \S\ref{sec:bitmasks}).

\subsubsection{The Data Model for Sky Locations}

Files of sky locations derived by \dt\ at the pixel-level (see \S\ref{sec:skies}) are available online in the {\tt dr9} directory in a similar way to the target files detailed in \S\ref{sec:btanddt}, with directory names that terminate with {\tt skies} rather than {\tt targets}\footnote{For example, the appropriate url for the Main Survey is \urlstub{dr9/1.1.1/skies/}.}. Files of supplemental sky locations derived by avoiding bright sources in {\em Gaia} (see \S\ref{sec:suppskies}) are available in {\tt gaiadr2} directories in a similar manner to what is outlined in \S\ref{sec:backup}. Directory names for supplemental skies, however, terminate in {\tt skies-supp} instead of in {\tt skies} or {\tt targets}\footnote{For example, Main Survey supplemental skies are at \urlstub{gaiadr2/1.1.1/skies-supp/}.}. 

Within the {\tt skies} ({\tt skies-supp}) directory, files have names of the form skies-hp-{\tt HPX}.fits (skies-supp-hp-{\tt HPX}.fits). Here, {\tt HPX} refers to the {\tt nside = 8} integer of the HEALPixel that contains the sky locations. As is the case for other \dt\ products, the adopted {\tt nside} is stored in the file header as {\tt FILENSID}. The {\tt skies} directory also contains an {\tt unpartitioned} sub-directory. This stores the original sky locations generated by {\tt desitarget} before applying the repartitioning code discussed in \S\ref{sec:skies}, i.e., {\tt unpartitioned} contains files assembled according to which {\em bricks} occupy each HEALPixel rather than by which {\em sky locations} occupy each HEALPixel. Columns in the {\tt skies} and {\tt supp-skies} files are either inherited from the Legacy Surveys sweep catalogs, are provided in Table~\ref{table:tscolumns}, or were introduced in \ref{sec:skies}. 

\begin{figure}[!t]
\begin{center}
\includegraphics[scale=0.365,trim=7 20 7 50,clip=true]{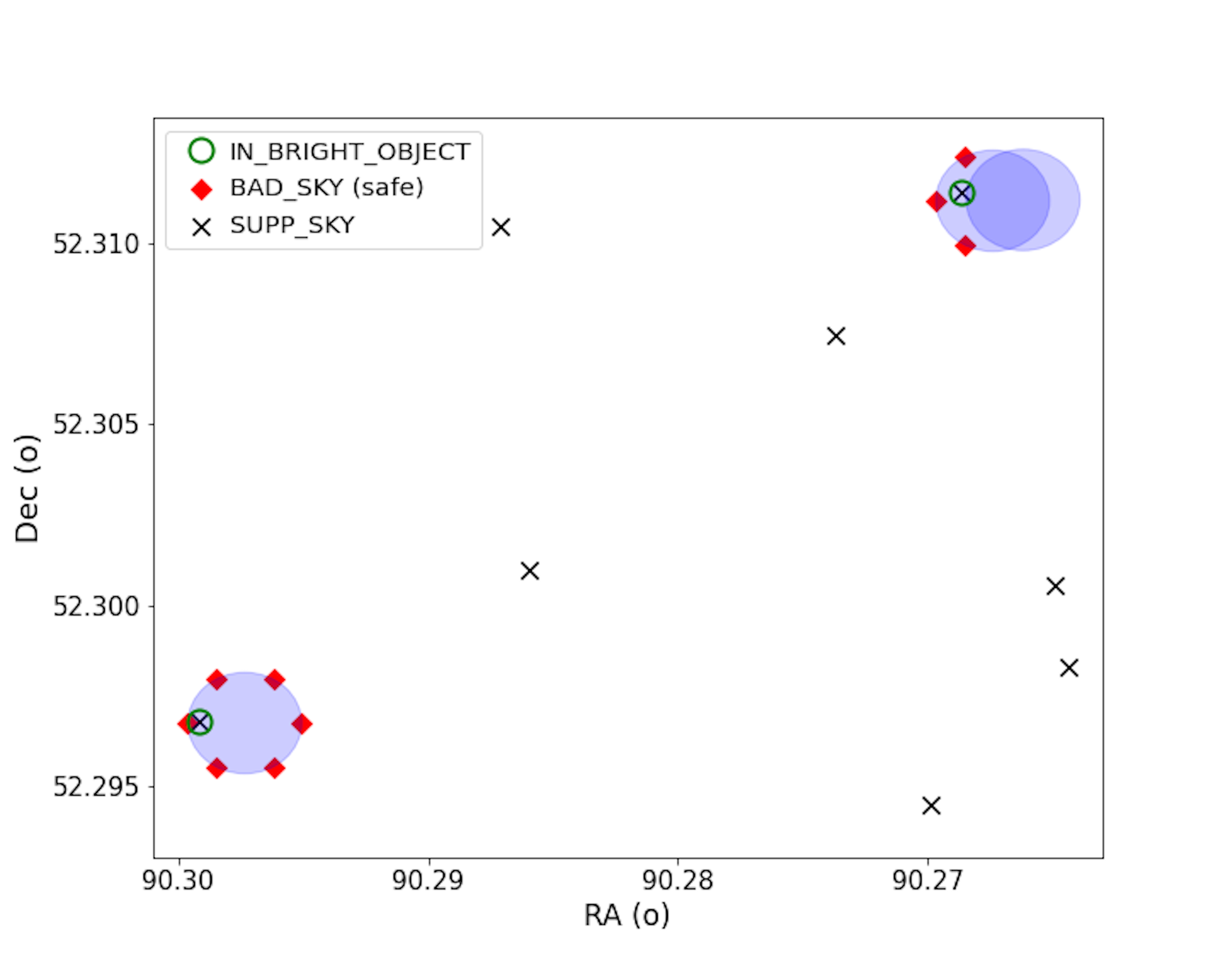}
\caption{Example blank sky locations. Crosses depict supplemental sky locations ({\tt SUPP\_SKY}; see \S\ref{sec:suppskies}). Large blue circles represent masks around bright stars, and red diamonds depict ``safe'' locations generated when a {\tt SUPP\_SKY} falls in a mask ({\tt BAD\_SKY}; see \S\ref{sec:mask}). {\tt SUPP\_SKY} locations in masks are not discarded, but have the {\tt IN\_BRIGHT\_OBJECT} bit set (green circles). The masks in the top right-corner demonstrate how safe locations that have {\em both} {\tt BAD\_SKY} and {\tt IN\_BRIGHT\_OBJECT} set {\em are} discarded: Safe ({\tt BAD\_SKY}) locations were generated around the left-hand star mask based on a {\tt SUPP\_SKY} in that mask, but half fell in the right-hand star mask and so were discarded.
\label{fig:sky}}
\end{center}
\end{figure}

\subsubsection{Masking and ``Safe'' Locations}
\label{sec:mask}

Locations assigned near very bright stars are potentially unusable for sky subtraction by the \desi\ spectroscopic pipeline. To help characterize such locations for SV and the Main Survey, the \dt\ code\footnote{Specifically the {\tt brightmask.make\_bright\_star\_mask} function.} was used to compile a ``bright star'' catalog using all {\em Gaia} DR2 sources, supplemented with sources from the Tycho 2 \citep{TYCHO2} catalog\footnote{\url{https://heasarc.gsfc.nasa.gov/W3Browse/all/tycho2.html}}. Any matches between {\em Gaia} and Tycho 2 within a radius of $10\arcsec$ are first removed from the Tycho 2 catalog to prevent duplication of sources. Then, {\em Gaia}-Tycho coordinate matches are performed, after first shifting the location of {\em Gaia} sources to the Tycho epoch using {\em Gaia} astrometric parameters. {\em Gaia} sources with missing proper motions are assigned information from URAT using the same routine as detailed for GFAs (see \S\ref{sec:gfas}). This catalog is then limited to only sources with a magnitude $<12$, where the magnitude that is used is {\em Gaia} $G$, Tycho $VT$, Tycho $HP$ and Tycho $BT$, in that order of preference---because $G$ is not measured for Tycho-only sources, $VT$ isn't measured for every Tycho source, etc.

A ``bright star mask'' is then constructed around all sources in this bright star catalog by shifting sources to an epoch of 2023.0 to roughly match the expected central time of the \desi\ survey. Any sky locations within 5\arcsec\ of this (circa 2023) bright star mask are assigned the bit {\tt IN\_BRIGHT\_OBJECT} from the {\tt desi\_mask} bitmask in the {\tt DESI\_TARGET} column (see \S\ref{sec:bitmasks}). Similarly, any sky locations within 10\arcsec\ of a bright star are assigned the {\tt NEAR\_BRIGHT\_OBJECT} bit.

Given the extensive sky coverage of the \desi\ survey, it is impossible to completely avoid placing fibers near bright stars. So, to assist with finding reasonable locations for sky subtraction in regions near bright stars, both the pixel-based and supplemental skies are augmented by additional ``safe'' locations. Any sky location that has {\tt IN\_BRIGHT\_OBJECT} set is supplemented with six additional, symmetrically placed locations (see Figure~\ref{fig:sky}) around the periphery of the circular mask for any individual star. This placement pattern corresponds to approximately 1 safe location per radius of the mask (with a margin of one extra location), which should be sufficient to ensure a safe location for the vast majority of fibers in the \desi\ focal plane. These safe locations all have the {\tt BAD\_SKY} bit set. They are also assigned the bits {\tt IN\_BRIGHT\_OBJECT} and {\tt NEAR\_BRIGHT\_OBJECT} using the bright star mask (as for any other sky location).

Finally, any pixel-based or supplemental sky locations that have both {\tt IN\_BRIGHT\_OBJECT} and {\tt BAD\_SKY} set (including ``safe'' locations) are removed completely from their respective \dt\ output files. This step ensures that any freshly generated safe locations aren't, themselves, inside the boundary of a bright star (again see Figure~\ref{fig:sky}). Sky locations with other combinations of the sky and bright object bits set are retained, but with all of the bits set in the {\tt DESI\_TARGET} column to help downstream code identify potentially problematic positions.

\begin{figure}[!t]
\begin{center}
\includegraphics[scale=0.395,trim=17 20 7 40,clip=true]{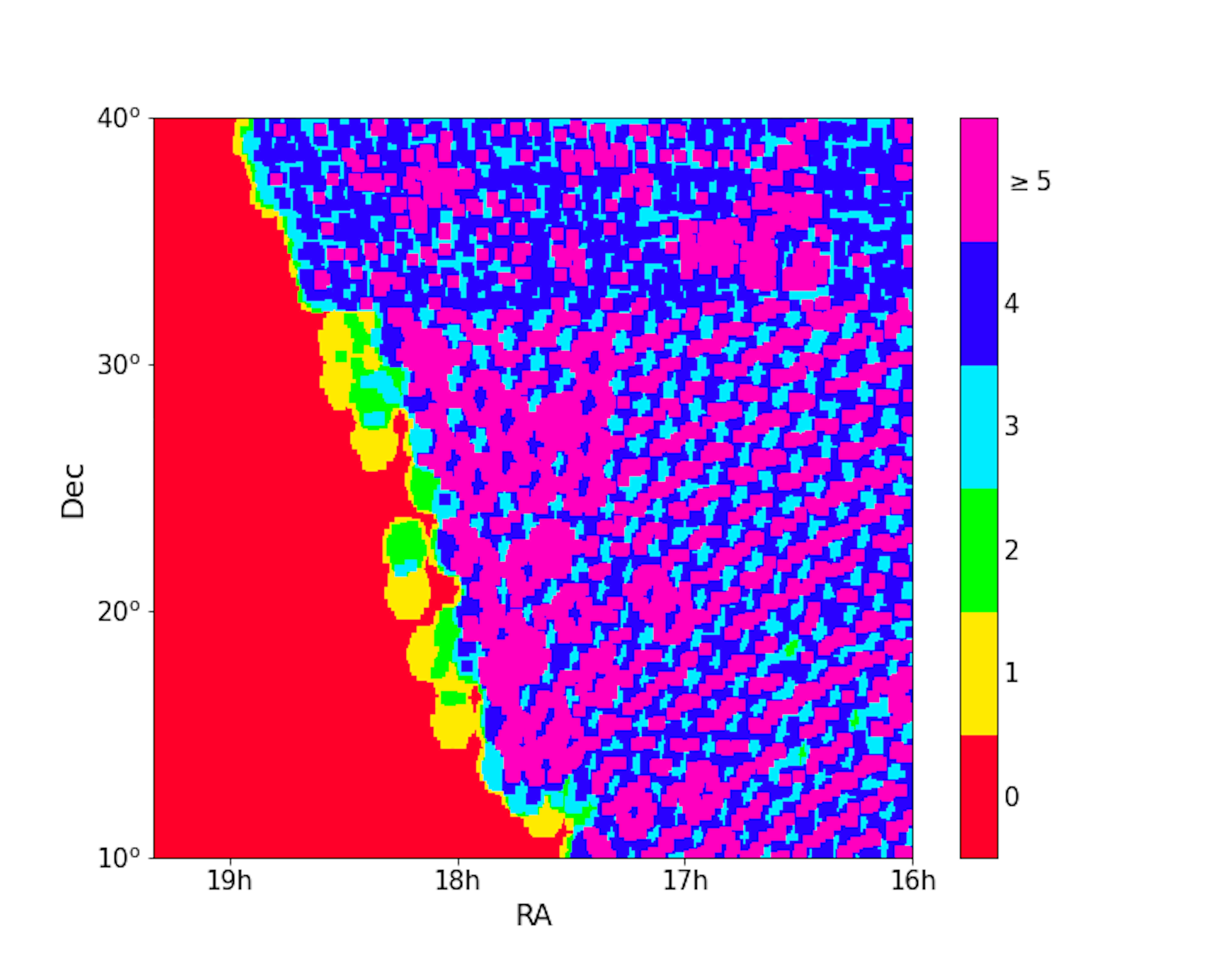}
\caption{An example use of the \dt\ ``allsky'' random catalogs. Colors distinguish different numbers of $g$-band observations in DR9 of the Legacy Surveys. For instance, areas colored green (pink) were covered by two ($\geq5$) observations. Areas that have zero observations because they are outside of the Legacy Surveys are immediately obvious, as is the lower coverage near the edge of the footprint. The division between BASS and MzLS at a declination of $\sim32.375^\circ$ is also apparent (see also \S\ref{sec:resolve}). It is trivial to use these random catalogs to characterize how much area in the Legacy Surveys match specified observational criteria.
\label{fig:nobs}}
\end{center}
\end{figure}

\subsection{Monte Carlo Sampling of the Target-selection Footprint}

To help facilitate large-scale structure analyses with \desi, the \dt\ pipeline produces Monte Carlo samples that attempt to recreate the angular selection function of the Legacy Surveys imaging that was used to select \desi\ targets. These ``random catalogs,'' and related products, are detailed in this section.

\begin{deluxetable*}{ccc}[t]
\tablecaption{Columns in the pixweight files}\label{table:pwcolumns}
\tablewidth{0pt}
\tablehead{
\colhead{Column Name} & 
\colhead{Python Format} &
\colhead{Description}
}
\startdata
{\tt HPXPIXEL}    & {\tt i4 } & HEALPixel integers at an input {\tt nside} (typically {\tt nside=256}). \\
{\tt FRACAREA}    & {\tt f4 } & Fraction of the HEALPixel with $\geq 1$ observation in any band.\tablenotemark{a} \\
{\tt STARDENS}    & {\tt f4 } & Density ($\mathrm{deg}^{-2}$) of stars in the HEALPixel from {\em Gaia} DR2 (see \S\ref{sec:pixweight}). \\
{\tt EBV}         & {\tt f4 } & Galactic dust extinction (E[B-V]).\tablenotemark{b} \\
{\tt PSFDEPTH\_X} & {\tt f4 } & $\frac{5}{\sqrt{{\tt PSFDEPTH\_X}}}$ gives the 5$\sigma$ flux-detection-limit in nanomaggies for a point source.\tablenotemark{b, c} \\
{\tt GALDEPTH\_X} & {\tt f4 } & As for {\tt PSFDEPTH\_X} but for a round, exponential galaxy of effective radius $0.45\arcsec$.\tablenotemark{b, c} \\
{\tt PSFSIZE\_X}  & {\tt f4 } & Weighted average PSF FWHM, in arcseconds.\tablenotemark{b, d} \\
{\tt FRACAREA\_X} & {\tt f4 } & Fraction of HEALPixel with $\geq 1$ observation in any band with MASKBITS=X.\tablenotemark{e} \\
{\tt \em{target name}} & {\tt f4 } & Target density in the HEALPixel for each target class bitname (see Table~\ref{table:tspapers}).\\
\enddata
\tablenotetext{a}{Derived from the {\tt NOBS\_G}, {\tt NOBS\_R}, {\tt NOBS\_Z} quantities in the random catalog.}
\tablenotetext{b}{This quantity is the median in the HEALPixel from the passed random catalog.}
\tablenotetext{c}{One each for optical and {\em WISE} bands (i.e.\ {\tt X} = {\tt G}, {\tt R}, {\tt Z}, {\tt W1}, {\tt W2}) where the {\em WISE} values are AB, not Vega.}
\tablenotetext{d}{One for each optical band (i.e.\ {\tt X} = {\tt G}, {\tt R}, {\tt Z}).}
\tablenotetext{e}{This is a bitwise OR, so, e.g., if X=7 then it's the fraction for $2^0\, |\, 2^1\, |\, 2^2$. {\tt MASKBITS} is described in the Legacy Surveys documentation at \url{https://www.legacysurvey.org/dr9/bitmasks/}.}
\end{deluxetable*}

\subsubsection{Random Catalogs}
\label{sec:randoms}

The main \dt\ module used to create random catalogs is called {\tt desitarget.randoms} and the main worker function is {\tt randoms.quantities\_at\_positions\_in\_a\_brick()}. Initially, a desired density of random points ($\mathrm{deg}^{-2}$) is specified\footnote{This density appears in the header of output random catalogs as {\tt DENSITY}.}. Then, for each brick in the Legacy Surveys, 
the desired density is multiplied by the ($\sim0.0623\,\mathrm{deg}^2$) brick area, with the resulting number being modified by a (random) draw from the Poisson distribution to produce a final number of points to be sampled in a given brick. Locations (in RA and Declination) are then generated in the brick according to that final number of points.

The \dt\ pipeline then queries the Legacy Surveys imaging at the pixel-level, by directly sampling information from the CCDs used to produce the Legacy Surveys. In general, the files used as inputs to \dt\ in this context are from the Legacy Surveys coadded stack files\footnote{The stacks are documented at \lsurl{dr9/files/\#image-stacks-region-coadd}.}. Each random location is converted to a pixel using the World Coordinate System information for a given brick-based stack. The quantities produced for the \desi\ random catalogs are described in the documentation for the Legacy Surveys\footnote{Specifically, under ``random catalogs'' at \lsurl{dr9/files/\#random-catalogs-randoms}.}. Flux-related columns are derived by integrating over an aperture of a specified radius, as described in \S\ref{sec:skies}. The chosen aperture radius is recorded as {\tt APRAD} in the header of an output random catalog, and is likely to always be 0.75\arcsec, corresponding to the radius of a \desi\ fiber.

In addition to the random catalogs described above, the \dt\ pipeline can produce randoms catalogs in bricks that lie {\em outside} of the Legacy Surveys footprint. Such catalogs only include a limited number of columns, as most of the informative quantities are not defined where imaging does not exist. The ``inside'' and ``outside'' catalogs can also be combined by \dt\ to produce all-sky catalogs, which are particularly useful for determining, e.g., the areal coverage of the Legacy Surveys (see Figure~\ref{fig:nobs}). Again, columns present in these additional catalogs are detailed in the Legacy Surveys documentation.

As for other files produced by \dt, random catalogs are available online at the urls specified in \S\ref{sec:btanddt}. For random catalogs, the directory names terminate with {\tt randoms} (instead of, e.g., {\tt targets}) and the most recent \dt\ version number used to run a substantial set of random catalogs is {\tt 0.49.0}\footnote{i.e.\ the catalogs are at \urlstub{dr9/0.49.0/randoms/}.}. Versions of the random catalogs that have and have not been resolved according to the criteria in \S\ref{sec:resolve} are available in the {\tt resolve} and {\tt noresolve} sub-directories. A {\tt randomsall} sub-directory may also exist, which contains a larger file created by concatenating multiple smaller random catalogs to increase sampling density, as outlined in an accompanying {\tt README} file.

Within the {\tt randoms/resolve} directory, the inside-the-Legacy-Surveys random catalogs have the form randoms-{\tt ISEED}-{\tt ISPLIT}.fits, the outside-the-Legacy-Surveys random catalogs have the form randoms-outside-{\tt ISEED}-{\tt ISPLIT}.fits and the combined random catalogs have the form randoms-allsky-{\tt ISEED}-{\tt ISPLIT}.fits. Here, {\tt ISEED} refers to the random seed used to produce the catalog and {\tt ISPLIT} is an integer specifying a particular subset of randoms; the catalogs are split into smaller files from a larger parent set (which is itself randomized) during processing to make them more manageable. The specific integer seed ({\tt ISEED}) used to produce a given file is stored in the file header as {\tt SEED}.

\subsubsection{Pixelized Weight Maps}
\label{sec:pixweight}

The \dt\ pipeline---specifically the {\tt desitarget.randoms.pixmap()} function---can be used to combine files of \desi\ targets with random catalogs in a convenient HEALPixel-based format to produce maps of important quantities derived from the Legacy Surveys imaging. The resulting ``pixweight'' files have been made publicly available  for both SV3\footnote{e.g., \urlstub{dr9/0.57.0/pixweight/sv3/resolve/bright/sv3pixweight-1-bright.fits}} and the Main Survey\footnote{e.g., \urlstub{dr9/1.1.1/pixweight/main/resolve/dark/pixweight-1-dark.fits}}. Maps of dark-time and bright-time targets are created separately, and so each release directory contains {\tt dark} and {\tt bright} sub-directories. The HEALPixel resolution of each pixweight file is {\tt nside=256}, which is stored in the header of the file as {\tt HPXNSIDE}. This resolution corresponds to a pixel area of $\sim0.0525\,\mathrm{deg}^{2}$, which approximates the size of a Legacy Surveys brick.

The pixweight maps contain the quantities detailed in Table~\ref{table:pwcolumns}, which are all easily derived from the random catalogs discussed in \S\ref{sec:randoms} (typically from the file in the {\tt randomsall} sub-directory) and the bright- and dark-time target files described in \S\ref{sec:btanddt}. One column that appears in Table~\ref{table:pwcolumns} that we have not previously described is {\tt STARDENS}, which contains a measure of the stellar density in a given HEALPixel. The {\tt STARDENS} quantity is derived from a count of sources in {\em Gaia} DR2 that have a {\em Gaia} $G$-band magnitude in the range $12 \leq G < 17$ and that are point-like. Point-like, here, is defined using {\em Gaia} quantities via:

\begin{equation}
 AEN == 0.~\mathrm{OR}~\log10(AEN) < 0.3G - 5.3
\end{equation}

\noindent where $AEN$ is the {\em Gaia} quantity {\tt ASTROMETRIC\_EXCESS\_NOISE} and $G$ is the {\em Gaia} $G$-band magnitude (i.e.\ {\tt PHOT\_G\_MEAN\_MAG}).

\begin{figure*}[!t] 
\begin{center}
\hbox{
\includegraphics[scale=0.55434,trim=7 7 7 7,clip=true]{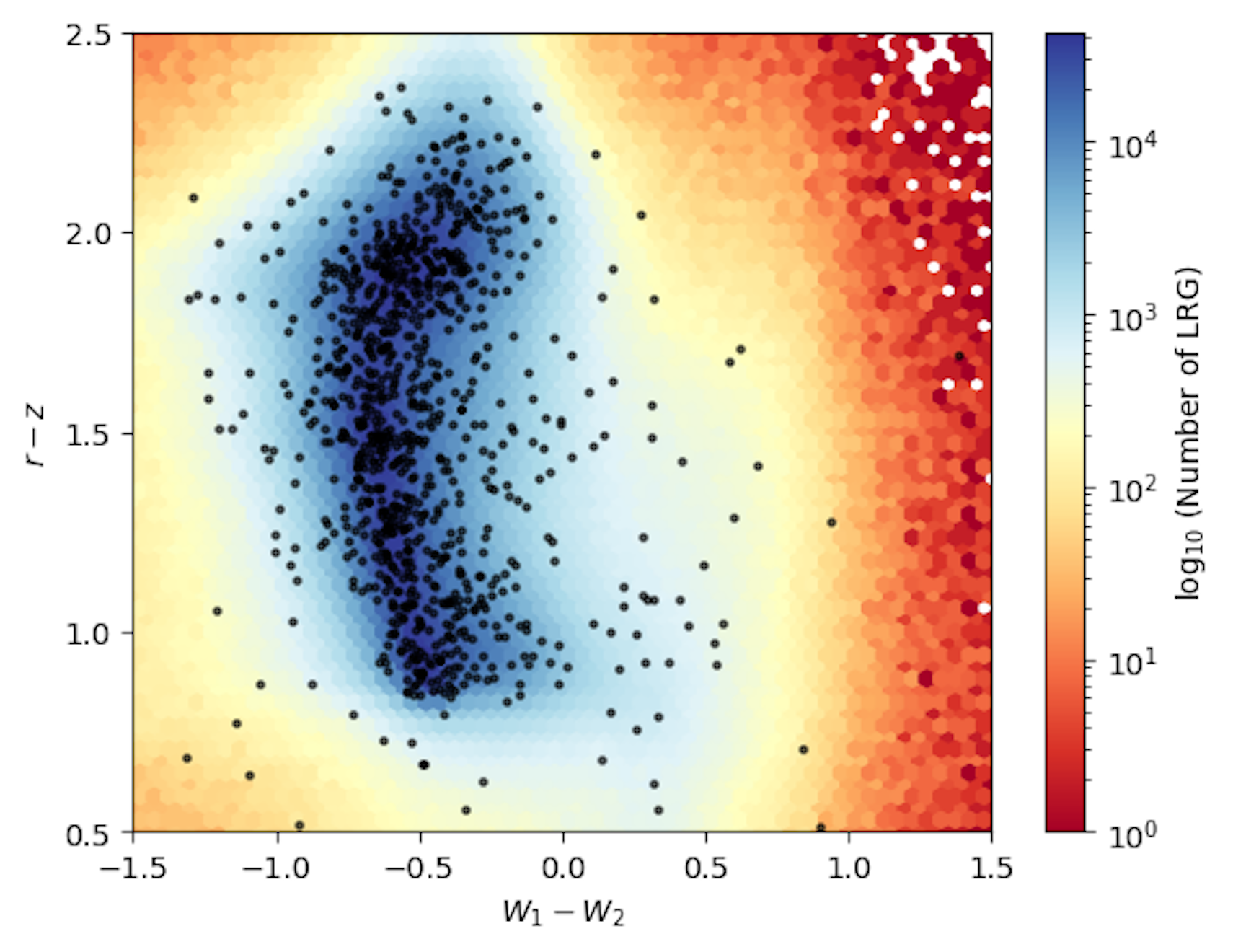}
\hspace{0.3cm}
\includegraphics[scale=0.5313,trim=17 8 45 8,clip=true]{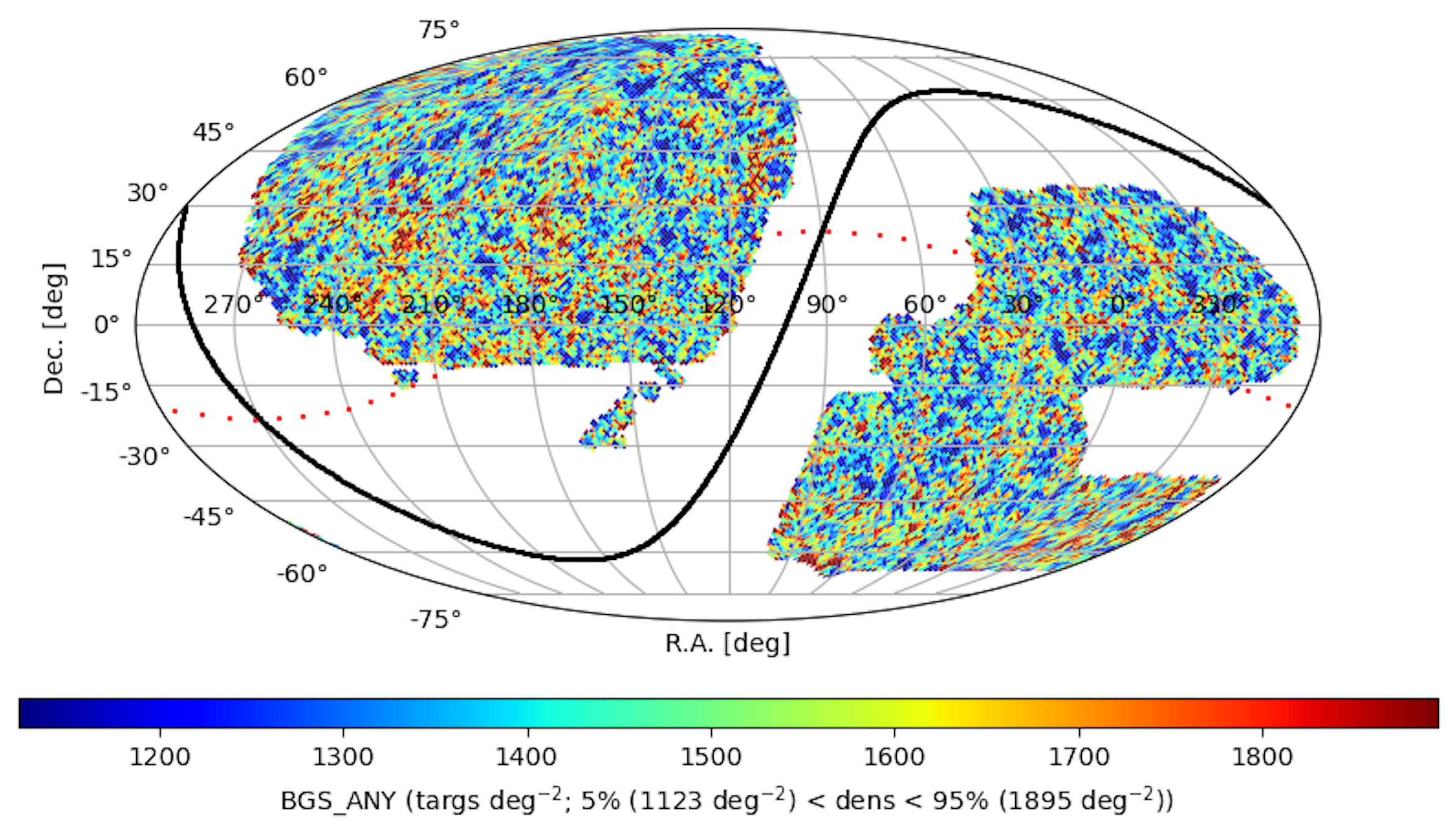}}
\caption{Example Quality Assurance (QA) diagnostic plots derived from running the \dt\ codebase on DR9 of the Legacy Surveys. {\em Left panel:} Number density contours for DESI Main Survey LRG (dark-time) targets in an optical color ($r - z$) compared to the 3.4 and 4.6\,$\mu$m bands ($W1 - W2$) from the {\em WISE} survey. {\em Right panel:} Density of all SV3 BGS targets in equatorial coordinates. The solid black line depicts the plane of our Galaxy and the red dotted line depicts the ecliptic.
\label{fig:QA}
}
\end{center}
\end{figure*} 

\subsubsection{Quality Assurance Web Pages}
\label{sec:QA}

The ``pixweight'' files described in \S\ref{sec:pixweight} can be used to derive useful plots of sky coverage for targets. In particular, the {\tt FRACAREA} quantity allows target densities to be correctly weighted by the actual areal coverage of the Legacy Surveys imaging. In addition, as the ``pixweight'' files collate information about observational properties of the imaging surveys, such as depth and local stellar density, they can be used to determine how target densities correlate with systematic effects \citep[see, e.g.][]{Cha22}.

The {\tt desitarget.QA} module can be used to combine the ``pixweight'' files from \S\ref{sec:pixweight} with the target files detailed in \S\ref{sec:btanddt} to create web pages of diagnostic plots for \desi\ targeting. The gateway to these web pages also includes a version of the target density matrices provided in Tables~\ref{table:targdensdark} and \ref{table:targdensbright}, which are trivial to derive from the ``pixweight'' files. 

These QA web pages have been made publicly available by the \desi\ collaboration for SV3\footnote{\urlstub{dr9/0.57.0/QA/sv3/}.} and the Main Survey\footnote{\urlstub{dr9/1.1.1/QA/main/}.}. At each url, there are individual sub-directories for bright-time and dark-time targets. Clicking on a sub-directory launches a page that includes target density matrices, plots of observational systematics, and a list of targeting bit-names (see, e.g., Table~\ref{table:tspapers}). Clicking on any bit-name produces a page with a number of diagnostic plots for that target class, including sky maps, color-color plots and trends in target density with observational systematics. Some examples of these diagnostic plots are included in Figure~\ref{fig:QA}.

\subsection{Secondary Target Classes}
\label{sec:secondary}

In addition to the dark-time targets, bright-time targets, backup targets,  calibration targets, and random catalogs discussed elsewhere in this section, the \desi\ survey incorporates additional targets to pursue science efforts beyond the primary \desi\ goals. These targets, which we refer to as ``secondary'' targets, are either conducted as dedicated campaigns or as ``filler'' programs to mop up spare fibers where no primary \desi\ targets are available. The dedicated campaigns include tertiary programs (c.f.\ \S\ref{sec:tertiary}) and some large Targets of Opportunity campaigns \citep[see \S\ref{sec:ToO} and, e.g.,][]{Pal21}.

\subsubsection{An Overview of Secondary Targets}

In \S\ref{sec:bitmasks} we discussed how secondary targets can be retrieved from the \desi\ target files (\S\ref{sec:btanddt}) by applying the {\tt scnd\_mask} bit-names to bit-values in columns that resemble {\tt SCND\_TARGET}. In \S\ref{sec:sectid} we detailed the specific, unique {\tt TARGETID} that \dt\ assigns to secondary targets. But, a full description of all of the scientific goals that underlie the \desi\ secondary target programs is beyond the scope of this paper and will be reserved for forthcoming \desi\ Data Release papers (e.g., DESI Collaboration et al.\ 2023, in preparation). We will, however, provide an overview of the process for selecting secondary targets here.

Every secondary program provides lists of targets which are preserved for posterity in a specific directory\footnote{Which is publicly available at \urlstubsec{secondary}.}. This ``secondary directory'' contains sub-directories for SV1, SV3, the Main Survey and some bespoke programs ({\tt sv1}, {\tt sv3}, {\tt main}, {\tt bespoke}; no secondary targets were assigned during SV2). Secondary targets added after the Main Survey may also be present in additional directories named {\tt main2}, {\tt main3}, etc. Each of these sub-directories includes; a file of notes about each secondary target class; a {\tt docs} directory of text files or Jupyter notebooks documenting a particular secondary program; an {\tt indata} directory of the targets supplied by a secondary program to be processed by \dt; and an {\tt outdata} directory containing files for bright-time ({\tt bright}) and dark-time ({\tt dark}) secondary targets that include the {\tt TARGETID}, {\tt DESI\_TARGET} and {\tt SCND\_TARGET} assigned by \dt\ (again see \S\ref{sec:bitmasks} and \S\ref{sec:sectid}). The {\tt outdata} directory contains one or more sub-directories, which correspond to the different versions of the \dt\ pipeline that were run to produce secondary targets. Each secondary targeting file is named {\tt \$bit-name.fits}, {\tt \$bit-name.txt} or {\tt \$bit-name.ipynb} where {\tt \$bit-name} is the specific bit-name assigned to a secondary program in the {\tt scnd\_mask} (see \S\ref{sec:bitmasks}).

Files of targets provided by a secondary program were required to include the columns {\tt RA}, {\tt DEC}, {\tt PMRA}, {\tt PMDEC}, {\tt REF\_EPOCH}, and {\tt OVERRIDE}. The first five of these columns have the same meaning as for the Legacy Surveys sweep catalogs\footnote{\lsurl{dr9/files/\#sweep-catalogs-region-sweep}.}. The {\tt OVERRIDE} column, which was required to be set to {\tt True} or {\tt False} for each secondary target for each program, is used by \dt\ to determine whether a secondary target should be {\em merged} with a primary target. When a secondary target {\em is} merged with a primary target by \dt\ (corresponding to {\tt OVERRIDE==False}), it inherits {\em all} of the properties of that primary target (such as the target's coordinates, observational priority, whether it will be observed in bright-time or dark-time, the primary {\tt TARGETID}, etc.). If, on the other hand, {\tt OVERRIDE} is set to {\tt True} for a target, then the target is free to be observed as defined by the secondary program but a new {\tt TARGETID} is generated for the secondary target (see \S\ref{sec:sectid}). Secondary and primary targets are merged by performing a coordinate match with a $1\arcsec$ separation\footnote{This separation is recorded in output \dt\ secondary files as {\tt SEP}.}.

One benefit to a secondary program of specifying {\tt OVERRIDE==False} for a secondary target that matches a primary target is that such targets are easier to track, because the same {\tt TARGETID} will be assigned for the secondary and primary target throughout the \desi\ program. This choice also guarantees standard processing for a secondary target (such as being automatically coadded by the spectroscopic pipeline in the same manner as for primary targets). But, a major drawback of providing targets with {\tt OVERRIDE==False} is that the target exactly resembles any matching \desi\ primary target, meaning that bespoke observational strategies---such as observing a target twenty times, or observing a match to a bright-time primary target in dark-time---cannot be adopted.

Secondary targets that specified {\tt OVERRIDE==False} are present in the standard dark-time and bright-time files detailed in \S\ref{sec:btanddt}. Whether a target is both a secondary and a primary target can be determined by checking if the {\tt SCND\_ANY} bit from the {\tt desi\_mask} is set in the {\tt DESI\_TARGET} column (see \S\ref{sec:bitmasks}). Secondary targets that specified {\tt OVERRIDE==True}\footnote{These targets are sometimes referred to as ``standalone'' secondaries in \desi\ parlance.} are written to their own monolithic files. These files are in the same root directory as detailed in \S\ref{sec:btanddt} for primary targets, except that instead of being stored in the {\tt resolve} directory they are in the {\tt secondary} directory. Bright-time standalone secondary targets are compiled in the {\tt secondary/bright/targets-bright-secondary.fits} file and dark-time standalone secondary targets are in the {\tt secondary/dark/targets-dark-secondary.fits} file.

Secondary targets that have {\tt OVERRIDE} set to {\tt False} aren't necessarily as carefully vetted as primary targets. So, the \dt\ code takes several precautions to ensure that \desi\ fibers aren't placed on very-bright {\tt OVERRIDE} targets that could contaminate the spectra of primary targets. First, any {\tt OVERRIDE} targets that lie inside of the bright star mask discussed in \S\ref{sec:mask} are removed completely from the output files of secondary targets. Second, when secondary targets are matched to primary targets (see above), they are also matched to the Legacy Surveys sweep catalogs\footnote{\lsurl{dr9/files/\#sweep-catalogs-region-sweep}.}. During this match, information from {\em Gaia} and the Legacy Surveys is added to the secondary targeting file, including flux and magnitude quantities ({\tt FLUX\_G}, {\tt FLUX\_R}, {\tt FLUX\_Z}, {\tt GAIA\_PHOT\_G\_MEAN\_MAG}, {\tt GAIA\_PHOT\_BP\_MEAN\_MAG}, {\tt GAIA\_PHOT\_RP\_MEAN\_MAG}). If an {\tt OVERRIDE} target is brighter than 16th magnitude in any of these {\em Gaia} or Legacy Surveys bands, it is removed completely from all of the output files of secondary targets.

\subsubsection{Is This Target Primary?}

The introduction of secondary programs and Targets of Opportunity somewhat complicate the provenance of a given target---particularly as secondary programs were allowed to be merged with primary programs if {\tt OVERRIDE} was set to {\tt False}. So, it is useful to provide a simple recipe for applying the information in \S\ref{sec:bitmasks} and \S\ref{sec:targetid} to determine whether a target is part of the primary \desi\ campaign or is more exotic. One possible recipe is:

\begin{itemize}

\item Check the {\tt DESI\_TARGET} column for a target, or the equivalent {\tt SVX\_DESI\_TARGET} column (see \S\ref{sec:bitmasks}).

\item If {\tt DESI\_TARGET} {\em only} has the bit SCND\_ANY set (i.e.\ if it is equal to $2^{62}$), then the target is either a secondary or tertiary target, or a ToO.

\begin{itemize}
   \item Otherwise the target is primary (consult \S\ref{sec:primary}).
\end{itemize}

\item Use the {\tt decode\_targetid()} utility (see \S\ref{sec:decode}) to determine {\tt release} for the target.

\begin{itemize}
    \item If {\tt release < 1000} then the target is a general secondary target (see Eqn.\,\ref{eqn:release} for the different meanings of {\tt release} and DESI Collaboration et al.\ 2023, in preparation, for a description of the secondary programs).
    \item If {\tt release == 8888} then the target is a tertiary target (consult \S\ref{sec:tertiary} and A.\ Raichoor et al.\ 2023, in preparation).
   \item If {\tt release == 9999} then the target is a ToO (consult \S\ref{sec:ToO} and E.\ Schlafly et al.\ 2023, in preparation).

\end{itemize}

\end{itemize}

\subsection{How to Run the \dt\ Pipeline}
\label{sec:tutorial}

Researchers may wish to fully recreate a suite of \dt\ output files in order to conduct large-scale-structure studies in the context of \desi\ or to inform target selection for future sky surveys. To help facilitate such endeavors, we have provided a tutorial on GitHub, in the form of a Jupyter Notebook, that includes a detailed overview of how each of the data products outlined in this section were produced for the \desi\ Main Survey using the \dt\ pipeline\footnote{See \url{https://github.com/desihub/desitarget/blob/2.5.0/doc/nb/how-to-run-target-selection-main-survey.ipynb}.}.

\section{Discussion: Important Known Issues}
\label{sec:discussion}

As with any evolving software, \dt\ underwent several bug-fixes during catalog-creation. In this section, we try to detail some of the most prominent issues and discrepancies that users of the DESI targeting files might encounter.

\subsection{Target Reproducibility}
\label{sec:targrep}

Perhaps the most important known issue is that output from the \dt\ code was not wholly reproducible throughout the Main Survey. This was due to the fact that although primary targets typically supersede secondary targets (see \S\ref{sec:secondary}) the \dt\ code was written to {\em allow Milky Way Survey targets to bump each other}, regardless of whether they were primary or secondary. Unfortunately, because some MWS targets have similar colors to primary targets---for instance, white dwarfs resemble some QSOs---and because MWS white dwarf targets had a relatively high observational priority, the \dt\ pipeline fostered occasional glitches where an MWS target could bump a primary target intended for large-scale structure analyses. Notably, in situations where multiple MWS targets matched a primary target during primary-secondary merging (\S\ref{sec:secondary}), {\em which} MWS target was chosen as the match was not ordered. So, on repeated execution of the \dt\ code a {\em different} MWS target could end up being merged with a particular primary target. The upshot of this bug was that, prior to version {\tt 1.1.1} of \dt, a small subset of bright-time targets, and very occasional dark-time targets, were selected differently each time \dt\ was run.

This MWS-prioritizing issue was noticed because a new, high-density, target class (called {\tt MWS\_FAINT}) was designed to be introduced for the first iteration of Main Survey targets, which were processed with version {\tt 1.0.0} of \dt. Due to an oversight, these {\tt MWS\_FAINT} targets were {\em not} included in the {\tt 1.0.0} target catalogs. When this oversight was corrected, and the target catalogs were updated, it became clear that a slightly different subset of targets had been selected, even when the {\tt MWS\_FAINT} targets were redacted. Any reproducibility bugs were subsequently fixed and new targets were processed with the {\tt MWS\_FAINT} selection once more omitted---resulting in the version {\tt 1.1.1} targets discussed throughout this paper.

While this bug-fix was being addressed, 157 Main Survey tiles were observed with version {\tt 1.0.0} targets before updating to version {\tt 1.1.1} of \dt\ and producing the bright-time and dark-time targets that will be used throughout the remainder of the \desi\ survey. Fortunately, the scope of the differences between which {\tt 1.0.0} targets {\em were} scheduled to be observed and which {\tt 1.1.1} targets {\em would have been} scheduled to be observed is minor. Across the 157 initial tiles, a total of 270 bright-time targets differed in a manner that could have produced discrepancies in fiber placement in the \desi\ focal plane\footnote{Note that the crucial point is not that 270 different targets {\em were} observed, merely that a maximum of 270 different targets {\em could have} been observed, potentially impacting the survey selection function for large-scale structure studies.}. Only {\em four} dark-time targets differed in such a way. Given that there were $\sim$785k fibers that could have been assigned fibers across 157 tiles, any reproducibility issues moving from version {\tt 1.0.0} to {\tt 1.1.1} targets near the start of the \desi\ Main Survey should have a negligible effect on large-scale-structure analyses.

\subsection{{\tt SUBPRIORITY} Reproducibility}

Assigning fibers to targets during early \desi\ Main Survey observations revealed an additional issue that could affect the overall reproducibility of the \desi\ survey. As noted in Table~\ref{table:tscolumns}, \dt\ adds a {\tt SUBPRIORITY} column to target files that is used to break ties in the event that the \desi\ fiber-assignment code has to choose between two targets that are otherwise equivalent in observational priority. A similar {\tt SUBPRIORITY} column is added by \dt\ to the pixel-based and supplemental sky locations discussed in \S\ref{sec:blankskies}. As of version {\tt 1.0.0} of the \dt\ pipeline, {\tt SUBPRIORITY} values were added as part of the function that writes files to disk. This meant that if an external routine read a target (or sky) file produced by \dt\ and then wrote it back out using the \dt\ writing utilities {\em then each {\tt SUBPRIORITY} would be overwritten with a new value}. 

Rewriting {\tt SUBPRIORITY} using the approach of version {\tt 1.0.0} of the \dt\ pipeline was not a bug, {\em per se}, but it proved problematic downstream of \dt. In particular, the software that assigns fibers to \desi\ targets (A.\ Raichoor et al.\ 2023, in preparation) uses the \dt\ writing routines to produce its own output data files. To guard against accidentally overwriting {\tt SUBPRIORITY} values throughout the Main Survey, the \dt\ pipeline was updated to only overwrite zero-valued {\tt SUBPRIORITY} entries---meaning that no values are accidentally overwritten when the \desi\ fiber assignment pipeline (or any other software) uses the \dt\ writing functions. Although this issue with {\tt SUBPRIORITY} was fixed before the Main Survey proceeded with version {\tt 1.1.1} targets, differences would persist in how fibers were assigned on the 157 Main Survey tiles that used targets compiled with version {\tt 1.0.0} of the \dt\ pipeline. To ameliorate any discrepancies, files were constructed that contained the mapping between {\tt SUBPRIORITY} and {\tt TARGETID} for the {\tt SUBPRIORITY} values that were initially assigned to version {\tt 1.0.0} Main Survey targets\footnote{These files are publicly available in the \urlstub{subpriority/fba-version-4.0.0/} directory.}. Versions of the \dt\ pipeline subsequent to {\tt 1.1.0} incorporate a command-line option that can use these mappings to enforce the values of {\tt SUBPRIORITY} used on the 157 Main Survey tiles that proceeded with version {\tt 1.0.0} targets\footnote{As detailed at \url{https://github.com/desihub/desitarget/blob/2.5.0/doc/nb/how-to-run-target-selection-main-survey.ipynb} in a Jupyter Notebook tutorial.}.

\subsection{Targets with {\tt RELEASE} of 9010 in SV1}

Early Survey Validation (i.e.\ SV1) took input information from the DR9 Legacy Surveys sweep catalogs before they were patched due to a data processing issue\footnote{see \lsurl{dr9/issues/\#bricks-that-were-processed-using-the-burst-buffer-at-nersc}.} that altered the imaging properties of some sources. The \dt\ pipeline switched to using the ``patched'' DR9 sweeps as of version {\tt 0.49.0} of the \dt\ code. Targets that might have different source properties because of this patching bug can be identified because they will have a {\tt RELEASE} of 9010 in files produced by versions of \dt\ prior to {\tt 0.49.0} but a {\tt RELEASE} of 9012 in later target files. Such targets will have similar coordinates but a different {\tt TARGETID}, both because the {\tt RELEASE} number changed for all such sources in the patched sweep catalogs {\em and} because the {\tt OBJID} changed for some such sources (see, e.g., Figure~\ref{fig:TARGETID}).

\section{Summary}
\label{sec:conclusions}

In this paper, we have provided a detailed overview of the \desi\ target selection pipeline. The exact algorithms used to classify specific subsets of \desi\ bright-time and dark-time targets are detailed in other papers \citep{elg, lrg, qso, mws, bgs} but aspects of how these targeting algorithms are {\em implemented} are crucial to understanding how to work with \desi\ data. This paper also details the approaches used to classify targets to help calibrate \desi\ spectroscopy, such as GFA targets and blank sky locations.

The \desi\ target selection pipeline, which is known as \dt, is publicly available on GitHub. As with any software product, some documentation of the \dt\ code and the resulting products will be described online, and in other papers related to the \desi\ end-to-end pipeline \citep[E.\ Schlafly et al.\ 2023, in preparation; A.\ Raichoor et al.\ 2023, in preparation;][S.\ Bailey et al.\ 2023, in preparation]{spec}. So, here, we have tried to focus on an overview of the most important concepts related to the technical aspects of \desi\ target selection.

The \dt\ pipeline sets bit-values in output data files in a number of different columns, which indicate whether a particular target meets specific selection criteria. The names of these columns resemble {\tt DESI\_TARGET} (for dark-time \desi\ targets), {\tt BGS\_TARGET} (for bright-time galaxies), {\tt MWS\_TARGET} (for Milky Way targets) and {\tt SCND\_TARGET} (for targets beyond the primary \desi\ campaigns). Values in these columns can be interpreted using bitmasks with names like {\tt desi\_mask}, {\tt bgs\_mask}, {\tt mws\_mask} and {\tt scnd\_mask}.

Another important output of the \dt\ pipeline is a unique {\tt TARGETID} that allows each target to be tracked throughout the \desi\ survey. This {\tt TARGETID} encodes additional, useful information about each \desi\ target, such as whether a target is selected from Legacy Surveys imaging or from {\em Gaia}, whether a target is a sky location or part of a random catalog, and whether a target is part of a secondary program or is a target-of-opportunity.

It is important to note that different versions of the \dt\ pipeline were used for different phases of the \desi\ survey, such as the Survey Validation phase---including the \desi\ One-Percent Survey---and the Main Survey phase. The evolving nature of \dt\ spawned multiple complexities in target selection that we have attempted to document in this paper. Examples of such technical details include (but are not limited to):

\begin{itemize}
    \item That sometimes {\em Gaia} EDR3 data, instead of {\em Gaia} DR2 data, is incorporated by \dt\ when processing certain target selection classes.
    \item That the URAT survey is used to help select GFA targets for sources that are missing astrometric information in {\em Gaia} DR2.
    \item That a bright-star-mask based on Tycho and {\em Gaia} is applied to all sky locations and secondary targets, and that secondary targets are further removed from any \desi\ program if they are too bright in {\em Gaia} or Legacy Surveys imaging.
    \item That targets processed using version {\tt 1.0.0} of the \dt\ pipeline were observed for the earliest part of the \desi\ Main Survey but that, for reproducibility reasons, \desi\ rapidly updated to observing targets processed with version {\tt 1.1.1} of the \dt\ pipeline.
\end{itemize}

The \dt\ codebase is now mostly stable as regards producing the static files used for \desi\ target selection. Further modifications to \dt\ might be expected in modules related to producing random catalogs for large-scale-structure analyses, and to converting static files to the dynamic files used to monitor the state of a \desi\ target as it is observed (see also E.\ Schlafly et al.\ 2023, in preparation).  Any updates will be documented in later works as the \desi\ survey progresses.

\vspace{\baselineskip}

ADM and JM were supported by the U.S.\ Department of Energy, Office of Science, Office of High Energy Physics, under Award Numbers DE-SC0019022 and DE-SC0020086. APC is supported by a Taiwan Ministry of Education Yushan Fellowship and Taiwan Ministry of Science and Technology grant 109-2112-M-007-011-MY3.

This research is supported by the Director, Office of Science, Office of High Energy Physics of the U.S.\ Department of Energy under Contract No.\ DE–AC02–05CH11231, and by the National Energy Research Scientific Computing Center, a DOE Office of Science User Facility under the same contract; additional support for DESI is provided by the U.S.\ National Science Foundation, Division of Astronomical Sciences under Contract No.\ AST-0950945 to the NSF’s National Optical-Infrared Astronomy Research Laboratory; the Science and Technologies Facilities Council of the United Kingdom; the Gordon and Betty Moore Foundation; the Heising-Simons Foundation; the French Alternative Energies and Atomic Energy Commission (CEA); the National Council of Science and Technology of Mexico (CONACYT); the Ministry of Science and Innovation of Spain (MICINN), and by the DESI Member Institutions: \url{https://www.desi.lbl.gov/collaborating-institutions}.

The DESI Legacy Imaging Surveys consist of three individual and complementary projects: the Dark Energy Camera Legacy Survey (DECaLS), the Beijing-Arizona Sky Survey (BASS), and the Mayall $z$-band Legacy Survey (MzLS). DECaLS, BASS and MzLS together include data obtained, respectively, at the Blanco telescope, Cerro Tololo Inter-American Observatory, NSF’s NOIRLab; the Bok telescope, Steward Observatory, University of Arizona; and the Mayall telescope, Kitt Peak National Observatory, NOIRLab. NOIRLab is operated by the Association of Universities for Research in Astronomy (AURA) under a cooperative agreement with the National Science Foundation. Pipeline processing and analyses of the data were supported by NOIRLab and the Lawrence Berkeley National Laboratory. Legacy Surveys also uses data products from the Near-Earth Object Wide-field Infrared Survey Explorer ({\em NEOWISE}), a project of the Jet Propulsion Laboratory/California Institute of Technology, funded by the National Aeronautics and Space Administration. Legacy Surveys was supported by: the Director, Office of Science, Office of High Energy Physics of the U.S.\ Department of Energy; the National Energy Research Scientific Computing Center, a DOE Office of Science User Facility; the U.S.\ National Science Foundation, Division of Astronomical Sciences; the National Astronomical Observatories of China, the Chinese Academy of Sciences and the Chinese National Natural Science Foundation. LBNL is managed by the Regents of the University of California under contract to the U.S. Department of Energy. The complete acknowledgments can be found at \url{https://www.legacysurvey.org/acknowledgment/}.

The authors are honored to be permitted to conduct scientific research on Iolkam Du’ag (Kitt Peak), a mountain with particular significance to the Tohono O’odham Nation.

This work has made use of data from the European Space Agency (ESA) mission
{\it Gaia} (\url{https://www.cosmos.esa.int/gaia}), processed by the {\it Gaia}
Data Processing and Analysis Consortium (DPAC,
\url{https://www.cosmos.esa.int/web/gaia/dpac/consortium}). Funding for the DPAC
has been provided by national institutions, in particular the institutions
participating in the {\it Gaia} Multilateral Agreement.

This research made use of Photutils, an Astropy package for
detection and photometry of astronomical sources \citep{photutils}. Some of the results in this paper have been derived using the healpy and HEALPix packages.


\section*{Data Availability}
 
Data generated or analyzed as part this work are publicly available at URLs included in the text of the paper. Data points and code to produce the figures from this paper are at \url{https://zenodo.org/record/7015786}.

\facilities{Blanco (DECam), Bok (90Prime), {\em Gaia}, Mayall (DESI, Mosaic-3), {\em NEOWISE}, {\em WISE}}

\software{{\tt astropy} \citep{astropy1, astropy2}, {\tt fitsio} (\url{https://github.com/esheldon/fitsio}), {\tt healpy} \citep{healpy}, {\tt matplotlib} \citep{matplotlib}, {\tt numpy} \citep{numpy}, {\tt photutils} \citep{photutils}, {\tt pyyaml} (\url{https://pyyaml.org/}), {\tt scipy} \citep{scipy}.}

\bibliography{desiTS}{}
\bibliographystyle{aasjournal}

\end{document}